\documentclass[twocolumn]{aastex62}
\usepackage{booktabs}
\usepackage{CJK}
\graphicspath{{./}{figures/}}
\received{August 15, 2018}
\revised{November 19, 2018}
\accepted{January 28, 2019}
\submitjournal{\apj}
\shorttitle{CO source structure of SPT-selected DSFGs}
\shortauthors{Dong et al.}
\begin{document}
\begin{CJK*}{UTF8}{gbsn}
\title{Source structure and molecular gas properties from high-resolution CO imaging of SPT-selected dusty star-forming galaxies}
\def\uf{Department of Astronomy, University of Florida, Gainesville, FL 32611, USA}
\def\ut{Department of Astronomy, University of Texas at Austin, 2515 Speedway Stop C1400, Austin, TX 78712, USA}
\def\chileone{Departamento de Ciencias Fisicas, Universidad Andres Bello, Fernandez Concha 700, Las Condes, Santiago, Chile}
\def\chiletwo{Millennium Institute of Astrophysics (MAS), Nuncio Monse\~nor Sotero Sanz 100, Providencia, Santiago, Chile}
\def\chile{N\'ucleo de Astronom\'{\i}a, Facultad de Ingenier\'{\i}a, Universidad Diego Portales, Av. Ej\'ercito 441, Santiago, Chile}
\def\france{Aix Marseille Univ., Centre National de la Recherche Scientifique, Laboratoire d'Astrophysique de Marseille, Marseille, France}
\def\eso{European Southern Observatory, Karl Schwarzschild Stra\ss e 2, 85748 Garching, Germany}
\def\nyc{Center for Computational Astrophysics, Flatiron Institute, 162 Fifth Avenue, New York, NY 10010, USA}
\def\stanford{Kavli Institute for Particle Astrophysics and Cosmology, Stanford University, Stanford, CA 94305, US}
\def\ua{Steward Observatory, University of Arizona, 933 North Cherry Avenue, Tucson, AZ 85721, USA}
\def\uci{Department of Physics and Astronomy, University of California, Irvine, CA 92697, USA}
\def\uiuc{Department of Astronomy, University of Illinois, 1002 West Green St., Urbana, IL 61801, USA}
\def\cfa{Harvard-Smithsonian Center for Astrophysics, 60 Garden Street, Cambridge, MA 02138, USA}
\def\mpi{Max-Planck-Institut f\"{u}r Radioastronomie, Auf dem H\"{u}gel 69 D-53121 Bonn, Germany}
\def\dalhousie{Department of Physics and Atmospheric Science, Dalhousie University, Halifax, Nova Scotia, Canada}
\correspondingauthor{Chenxing Dong (董辰兴)}
\email{dcx@ufl.edu}
\author[0000-0002-5823-0349]{Chenxing Dong (董辰兴)}
\affiliation{\uf}
\author[0000-0003-3256-5615]{Justin~S.~Spilker}
\affiliation{\ut}
\author[0000-0002-0933-8601]{Anthony~H.~Gonzalez}
\affiliation{\uf}
\author{Yordanka~Apostolovski}
\affiliation{\chileone}
\affiliation{\chiletwo}
\author[0000-0002-6290-3198]{Manuel~Aravena}
\affiliation{\chile}
\author{Matthieu~B\'ethermin}
\affiliation{\france}
\author{Scott~C.~Chapman}
\affiliation{\dalhousie}
\author[0000-0002-3805-0789]{\begin{CJK*}{UTF8}{bsmi}Chian-Chou~Chen (陳建州)\end{CJK*}\kern-0.35em}
\affiliation{\eso}
\author[0000-0003-4073-3236]{Christopher~C.~Hayward}
\affiliation{\nyc}
\author[0000-0002-8669-5733]{Yashar~D.~Hezaveh}
\affiliation{\stanford}
\author[0000-0002-4208-3532]{Katrina~C.~Litke}
\affiliation{\ua}
\author[0000-0003-4178-0800]{Jingzhe~Ma (马京哲)}
\affiliation{\uci}
\author[0000-0002-2367-1080]{Daniel~P.~Marrone}
\affiliation{\ua}
\author[0000-0002-5153-0920]{Warren~R.~Morningstar}
\affiliation{\stanford}
\author{Kedar~A.~Phadke}
\affiliation{\uiuc}
\author{Cassie~A.~Reuter}
\affiliation{\uiuc}
\author{Jarugula~Sreevani}
\affiliation{\uiuc}
\author[0000-0002-2718-9996]{Antony~A.~Stark}
\affiliation{\cfa}
\author[0000-0001-7192-3871]{Joaquin~D.~Vieira}
\affiliation{\uiuc}
\author[0000-0003-4678-3939]{Axel~Wei{\ss}}
\affiliation{\mpi}
\begin{abstract}
We present Atacama Large Millimeter/submillimeter Array (ALMA) observations of high-J CO lines ($J_\mathrm{up}=6$, 7, 8) and associated dust continuum towards five strongly lensed, dusty, star-forming galaxies (DSFGs) at redshift $z = 2.7$--5.7. These galaxies, discovered in the South Pole Telescope survey, are observed at $0.2''$--$0.4''$ resolution with ALMA.
Our high-resolution imaging coupled with the lensing magnification provides a measurement of the structure and kinematics of molecular gas in the background galaxies with spatial resolutions down to kiloparsec scales.
We derive visibility-based lens models for each galaxy, accurately reproducing observations of four of the galaxies. 
Of these four targets, three show clear velocity gradients, of which two are likely rotating disks.
We find that the reconstructed region of CO emission is less concentrated than the region emitting dust continuum even for the moderate-excitation CO lines, 
similar to what has been seen in the literature for lower-excitation transitions.
We find that the lensing magnification of a given source can vary by 20--50\% across the line profile, between the continuum and line, and between different CO transitions.
We apply Large Velocity Gradient (LVG) modeling using apparent and intrinsic line ratios between lower-J and high-J CO lines.
Ignoring these magnification variations can bias the estimate of physical properties of interstellar medium of the galaxies.
The magnitude of the bias varies from galaxy to galaxy and is not necessarily predictable without high resolution observations.
\end{abstract}
\keywords{
galaxies: high-redshift ---  
galaxies: ISM --- 
gravitational lensing: strong --- 
ISM: molecules}

\section{Introduction}
Dusty star-forming galaxies (DSFGs) contribute a significant fraction of the total star formation at high redshifts ($z>2$, see e.g. \citealt{Smail02,Barger12,Swinbank14,Smith17}), and host the most intense star formation in the universe, with rates up to thousands of solar masses per year \citep[e.g.][]{Hughes98,Chapman05,Casey14}.
Although rapid star formation is ongoing in these galaxies, the bright ultraviolet (UV) continuum from massive, young stars is obscured by dust. The interstellar dust absorbs and reprocesses the UV-optical light, radiating at far-infrared (FIR) and submillimeter wavelengths, rendering these galaxies optically very faint.
High redshift DSFGs were first detected at 850\ $\mu$m with very low spatial resolution ($>10''$, \citealt{Smail97,Barger98,Hughes98}). High resolution submillimeter detections required long integration times and sensitive interferometers to achieve spatial resolutions comparable to the typical size ($\sim$1\,kpc) of the star-forming regions \citep{Bussmann15,Simpson15,Spilker16}. 

For DSFG studies, understanding the molecular gas in the galaxies is important, as it is the fuel supply for continued star formation. Carbon monoxide (CO) is the second-most common molecule in the interstellar medium (ISM), after molecular hydrogen (H$_2$). 
It is the most commonly used tracer of molecular gas (i.e. H$_2$ gas) in the ISM. 
The intensity of different CO transitions depend on the physical conditions of the molecular gas in the ISM. In general, warmer and denser gas allows the CO molecules to occupy higher excitation levels.
The CO lines from lower excited levels (especially the ground state transition CO $J=1$-0) trace virtually all molecular gas
and are generally used to estimate the total mass of molecular gas \citep[e.g.][]{Bolatto13} and to analyze the star formation rate as a function of molecular mass of the system \citep[e.g.][]{Daddi10,Genzel10}. 
The higher-J CO lines are mainly excited by collisions with H$_2$ in warm, dense gas tracing active star forming regions.
Using multiple CO lines with different excitation levels, one can constrain the physical conditions of the observed system by applying radiative transfer modeling, such as the commonly-used Large Velocity Gradient (LVG) approximation \citep[e.g.][]{cradex}.

Gravitational lensing provides a means of increasing the efficacy of galaxy studies at high redshift ($z>1$).
It provides higher effective spatial resolution and can be used to study the sources in greater details than unmagnified ones \citep[e.g.,][]{Swinbank10,Spilker14}.
Recently, several large extragalactic surveys have detected a large population of these strongly lensed galaxies, e.g. the {\it Herschel} Astrophysical Terahertz Large Area Survey \citep[H-ATLAS;][]{Eales10,Negrello10,Negrello17}, the {\it Herschel} Multi-tiered Extragalactic Survey \citep[HerMES;][]{Oliver12,Bussmann15,Asboth16}, the Atacama Cosmology Telescope \citep[ACT;][]{Marsden14}, and the South Pole Telescope \citep[SPT;][]{Carlstrom11,Vieira13}.
SPT has discovered roughly 100 gravitationally lensed high-redshift DSFGs \citep{Vieira10,Hezaveh13,Spilker16}.
Subsequent studies find the sample has a median redshift of $\langle z\rangle=3.9$ \citep{Vieira13, Strandet16} with the highest redshift source at $z\sim7$ \citep{Strandet17,Marrone18}. 
\cite{Aravena13,Aravena16} conducted a survey of CO $J=1$--0 and $J=2$--1 line emission in 17 of these galaxies and find the gas masses in the range $(0.5$--$11)\times10^{10}\ M_\odot$,  gas depletion timescales $t_\mathrm{dep} < 200\ \mathrm{Myr}$, and CO to gas mass conversion factor $\alpha_\mathrm{CO}$  in the range $0.4$--$1.8\ M_\odot\ (\mathrm{K\ km\ s}^{-1}\ \mathrm{pc}^2)^{-1}$. These studies allow for the basic physical properties of these galaxies to be determined. 
More detailed studies on the intrinsic sizes, luminosities, and dynamics of these galaxies require lens modeling to account for the gravitational magnification and distortion \citep{Spilker16}.
If one can properly model the foreground lens, lensing in principle allows higher effective spatial resolution than can otherwise be obtained. It also presents new complications, because galaxies are extended sources and different regions of the galaxies can be magnified by different amounts depending on the lensing geometry and location relative to the lensing caustics (lines of theoretically infinite magnification).
This differential magnification can skew observed flux and line ratios if the observed image is analyzed directly without performing lens modeling, which might then introduce a bias in other derived quantities \citep{Hezaveh12,Serjeant12}. 
The solution is to be able to model as many of the relevant galaxy components as possible.

\cite{Hezaveh13} introduced a parameterized and visibility-based lens modeling approach and derived intrinsic sizes, FIR luminosities and star formation surface densities of four of the SPT sources. 
\cite{Spilker16} conducted similar modeling on 47 of these DSFGs with ALMA 870 $\mu$m continuum data and found that the median magnification of the SPT sample is $\sim$6, with a source size distribution similar to unlensed DSFGs.
\cite{Spilker15} applied the models to low-J CO lines of two galaxies and found that the 870 $\mu$m dust contiuum is more concentrated than the region traced by low-J CO.
The relatively larger CO sizes compared with dust have also been found in other unlensed DSFGs \citep[e.g.,][]{Tadaki17,Chen17}.
At high redshifts, high-J CO lines (up to $J_\mathrm{up}\sim 8$) are more accessible than the low-frequency ground-state transition, because they are brighter \citep[e.g.][]{Rosenberg15} and are redshifted into transparent atmospheric windows.
Inferences about the size distribution of DSFGs depend on the tracer observed. Thus, it is important to understand how the apparent sizes of different CO transitions and the dust continuum vary.

In this paper, we study five strongly-lensed DSFGs using high-J CO lines (CO(8-7), CO(7-6), or CO(6-5)), and dust continuum  near the frequencies of the corresponding line emission.
The sources were targeted as part of an ALMA Cycle 2 program to find dark matter sub-halos in the foreground lensing galaxies (Morningstar et al., in prep.), but the focus of this paper is the source structure of the background lensed galaxies in the sample.
With the modest resolution ($0.2''$--$0.4''$) of our CO maps, we can extend the size comparison of CO versus dust to high-J CO lines, and can also study in detail differential magnification and physical properties of the galaxies.

The structure of this paper is as follows.
In Section \ref{sec:obs} we describe the ALMA Cycle 2 high-J CO and dust continuum observations and other supporting data.
In Section \ref{sec:len} we describe the visibility-based lens modeling technique used in this study.
Analyses and results on the sizes and magnifications of the sources, and a demonstration of the effects of differential magnification on radiative transfer modeling are discussed in Section \ref{sec:ana}.
We present our conclusions in Section \ref{sec:con}.
Throughout this work, we assume a flat Planck $\Lambda$CDM cosmology, where $h = 0.677$, $\Omega_m = 0.308$, and $\Omega_\Lambda = 0.691$  \citep{Planck16}.

\section{Observations}\label{sec:obs}
We primarily use ALMA Cycle 2 (2013.1.00880.S, PI: Y.~Hezaveh) observations for this study, augmented by lower-J CO data from the Australia Telescope Compact Array (ATCA) and additional mid-J CO transitions at low spatial resolution from other cycles of ALMA.

\subsection{ALMA Cycle 2}
We observed five galaxies, SPT0346-52, SPT0529-54, SPT0532-50, SPT2134-50 and SPT2147-50, with the ALMA 12m array on 28-29 June 2015 and 5 August 2015, with integration times of 34 to 44 minutes each. 
The number of antennas varied from 37 to 41.
The spatial resolution reaches $0.2''$ to $0.4''$, corresponding to 1.7 to 2.9 kpc in the absence of lensing, and up to 2.5 times better than the observations presented in \cite{Spilker16}.
Basic details of the observations are given in Table \ref{tab:galainfo}.

\begin{table*}
   \centering
   \caption{Observations} 
   \begin{tabular}{@{} cccccccccccc @{}}
      \toprule
      Name & R.A. & Dec. & Redshift & CO transition & $\nu_\mathrm{rest,CO}$\tablenotemark{a}   & Integ.\tablenotemark{b}   & Res.\tablenotemark{c}    & $\sigma_\mathrm{line}$\tablenotemark{d}   & $\sigma_\mathrm{cont}$\tablenotemark{e}\\
      &(J2000)&(J2000)&&&(GHz)&(min)&(arcsec) &(mJy/beam)&(mJy/beam)\\
      \midrule
      SPT0346-52 & 03:46:41.09 & -52:05:02.2 & 5.656 & 8-7 & 921.8  & 44 & 0.39  & 0.62 & 0.015 \\
      SPT0529-54 & 05:29:03.09 & -54:36:40.0 & 3.369 & 6-5 & 691.5  & 38 & 0.38  & 0.77 & 0.020 \\
      SPT0532-50 & 05:32:51.04 & -50:47:07.5 & 3.399 & 6-5 & 691.5  & 37 & 0.34  & 0.76 & 0.020 \\
      SPT2134-50 & 21:34:03.34 & -50:13:25.1 & 2.780 & 7-6 & 808.0  & 35 & 0.22  & 0.59 & 0.019 \\
      SPT2147-50 & 21:47:19.05 & -50:35:54.0 & 3.760 & 6-5 & 691.5  & 43 & 0.38  & 0.57 & 0.014 \\
      \bottomrule
   \end{tabular}
   \tablecomments{\tablenotemark{a}CO rest-frame frequency. \tablenotemark{b}On-source integration time. \tablenotemark{c}Angular resolution. \tablenotemark{d}Line sensitivity per 10 km~s$^{-1}$. \tablenotemark{e}Continuum sensitivity. 
   }
   \label{tab:galainfo}
\end{table*}

For each source, we observe one high-J CO line, with a correlator spectral resolution of 3.9 MHz and a bandwidth of 1.875 GHz. We observe three spectral windows of continuum with 128 channels each within 14 GHz from the line frequency with total bandwidth 6 GHz and usable bandwidth 5.625 GHz.

\begin{figure*}
\epsscale{1.15}
\plotone{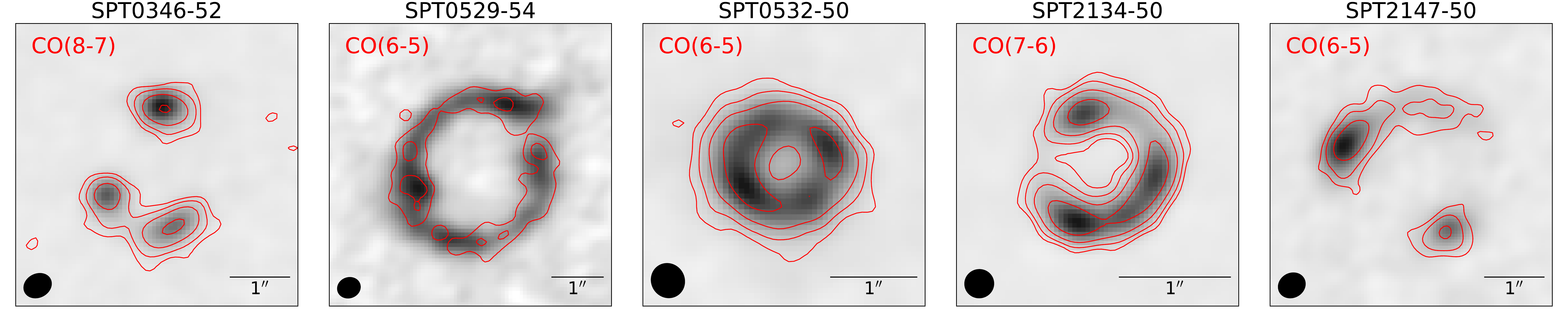}
\caption{ALMA Cycle 2 images of the five lensed galaxies studied here. The red contours indicate the CO line emission, while the greyscale corresponds to the dust continuum near the rest frequency of the targeted CO transition. The contours start at $3\sigma$ and increment by factors of $2^n$. The synthesized beams are indicated in the lower left corners and the $1''$ scale bars are indicated in the lower right corners.
\label{fig:img}}
\end{figure*}

All data was reduced using the standard ALMA Cycle 2 pipeline, with manual inspection of the quality of the reduction.
All lens analysis is performed using the interferometric visibilities (see Section \ref{sec:len}). For imaging, we use the \texttt{CASA} \citep{casa} \texttt{CLEAN} task with natural weighting to generate both the continuum image and the line data cube. 
For the line images, we subtract the continuum from the line data by fitting a linear polynomial to the continuum spectral window visibilities and applying that solution to the line spectral window. 
We also make integrated images of the CO lines by averaging the $uv$ data in frequency to create a single, very wide channel,
with width 800\,km\,s$^{-1}$ for SPT0346-52, SPT0532-50 and SPT2134-50, 700\,km\,s$^{-1}$ for SPT0529-54 and 500\,km\,s$^{-1}$ for SPT2147-50. 
The single-channel widths capture the bulk ($>80\%$) of the total line flux in each case.
These line images are shown as contours overlaid on the continuum images in Figure \ref{fig:img}. 

\begin{figure*}
\epsscale{1.2}
\plotone{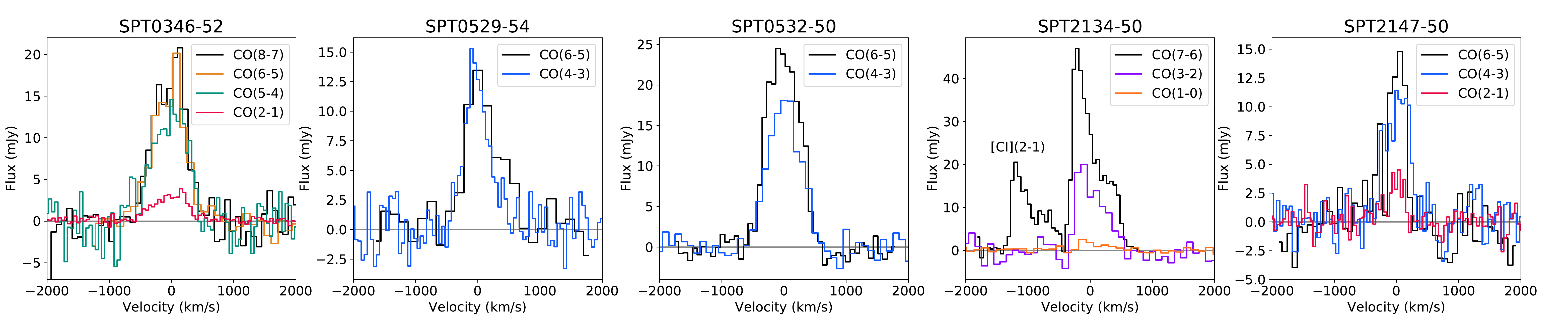}
\caption{Spectra extracted from the ALMA and ATCA data. The black spectra are from our ALMA high-J CO data. The colored spectra are from ALMA Cycle 0 mid-J data and ATCA low-J data. For SPT2134-50, the [CI](2-1) line is also detected at $-1200$ km~s$^{-1}$. Generally, as expected, the higher-J CO lines are brighter, while the FWHM remains nearly constant.
\label{fig:spec}}
\end{figure*}

We extract spectra using $1''$ to $2.5''$ radius apertures and 38\,MHz channel width for all sources except SPT0529-54. For this source, we use 76\,MHz channels because the data have a much lower signal-to-noise ratio (SNR). These spectral resolutions correspond to 50 to 150 km~s$^{-1}$ (See Table \ref{tab:galainfo}). The extracted spectra are shown in Figure \ref{fig:spec}. For the galaxy SPT2134-50, we also detect the [CI](2-1) line, which shows a similar profile to the adjacent CO(7-6) line. 

To estimate the line properties, we fit Gaussian functions to the spectra, except for SPT2134-50, which has a non-gaussian and asymmetric line profile.
For this source, we estimate the line flux by integrating the flux channel by channel and directly measuring the FWHM. We estimate the uncertainties by adding gaussian noise with a level from the actual spectrum to the spectrum and repeat the procedure for getting the flux and FWHM 1000 times.
The estimated line widths and fluxes are summarized in Table \ref{tab:lineimgflux} and are used to perform LVG radiative transfer modeling (see Section \ref{sec:lvg}).
The continuum flux densities are also estimated and are listed in Table \ref{tab:sizecontfluxmass}, to be compared with the intrinsic flux densities derived from lens modeling.

\begin{table*}
\centering
   \caption{Observed apparent line fluxes (without magnification correction)}
   \begin{tabular}{cclrrcc}
\toprule
Source & Line & FWHM & $S\mathrm{d}v$ & $L^\prime_\mathrm{CO}$ & Reference \\
  &   & (km s$^{-1}$) & (Jy km s$^{-1}$) & ($10^{10}$ K km s$^{-1}$ pc$^2$) &   \\
\midrule
SPT0346-52 & CO(2-1) & 613  $\pm$ 30 & 2.1  $\pm$ 0.1 & 51.6  $\pm$ 3.6 & [1] \\
 & CO(5-4) & 479  $\pm$ 209 & 8.4  $\pm$ 0.7 & 37.5  $\pm$ 3.2 & [2] \\
 & CO(6-5) & 619  $\pm$ 39 & 11.3  $\pm$ 0.9 & 35.0  $\pm$ 2.9 & [3] \\
 & CO(8-7) & 630  $\pm$ 46 & 13.1  $\pm$ 1.2 & 22.9  $\pm$ 2.2 & [4] \\
SPT0529-54 & CO(4-3) & 485  $\pm$ 108 & 6.7  $\pm$ 0.5 & 21.0  $\pm$ 1.6 & [2] \\
 & CO(6-5) & 585  $\pm$ 68 & 7.6  $\pm$ 1.1 & 10.6  $\pm$ 1.6 & [4] \\
SPT0532-50 & CO(4-3) & 416  $\pm$ 144 & 11.2  $\pm$ 0.6 & 35.6  $\pm$ 1.8 & [2] \\
 & CO(6-5) & 625  $\pm$ 22 & 17.2  $\pm$ 0.8 & 24.4  $\pm$ 1.1 & [4] \\
SPT2134-50 & CO(1-0) & 469  $\pm$ 180 & 1.0  $\pm$ 0.2 & 34.8  $\pm$ 6.3 & [1] \\
 & CO(3-2) & 522  $\pm$ 143 & 9.2  $\pm$ 0.8 & 37.3  $\pm$ 3.4 & [2] \\
 & CO(7-6) & 274  $\pm$ 22 & 21.6  $\pm$ 0.3 & 16.1  $\pm$ 0.3 & [4] \\
SPT2147-50 & CO(2-1) & 290  $\pm$ 52 & 1.2  $\pm$ 0.2 & 16.0  $\pm$ 3.2 & [1] \\
 & CO(4-3) & 378  $\pm$ 137 & 5.3  $\pm$ 0.5 & 19.8  $\pm$ 1.8 & [2] \\
 & CO(6-5) & 364  $\pm$ 47 & 5.6  $\pm$ 1.0 & 9.3  $\pm$ 1.6 & [4] \\
\bottomrule
   \end{tabular}
   \tablecomments{References: [1] \cite{Aravena16}, [2] \cite{weiss13}, [3] Apostolovski et al. in prep., [4] This work.}
   \label{tab:lineimgflux}
\end{table*}

\subsection{Additional ALMA and ATCA data}
We use the ALMA Cycle 0 (2011.0.00957.S, PI: A.~Wei\ss) low resolution 3\,mm data which were originally used to derive redshifts for these sources. For each source the data contains one or two mid-J CO transitions from CO(3-2) to CO(6-5).
Details of the data and observations are given in \cite{weiss13}.
For the CO(6-5) line of SPT0346-52, we have obtained additional ALMA data (2015.1.00117.S, PI: M.~Aravena) with higher SNR and better flux calibration, so we use the newer line flux. Details of these data and observations will be given in Apostolovski et al. in prep.
We also use 4--5'' resolution CO(1-0) and CO(2-1) data from ATCA 
for SPT0346-52 (CO(2-1)), SPT2134-50 (CO(1-0)) and SPT2147-50 (CO(2-1)) \citep{Aravena16}. 
In Figure \ref{fig:spec}, we overplot the lower-J CO spectra from these additional ALMA and ATCA data. 
Spectra of different transitions for a given galaxy generally have similar shape and FWHM, unlike some observations of unlensed DSFGs \citep{Ivison11}.
For SPT0346-52 and SPT2147-50, $\sim$0.5'' resolution observations of the same low-J CO transitions are also available, and we perform lens modeling for these higher-resolution data and use the output for LVG modeling (\citealt{Spilker15}, Apostolovski et al. in prep.).

\section{Lens modeling}\label{sec:len}
We use the parameterized lens modeling code \texttt{visilens} described in \cite{Spilker16}, following \cite{Hezaveh13}. Rather than modeling the images reconstructed from radio interferometer data, we perform our analysis directly in the visibility plane. With this technique we can better understand the measurement and its noise, avoiding the correlated noise inherent to interferometric images, and can account for residual calibration errors.

Within the model, the lens is described by one or more Singular Isothermal Ellipsoids (SIEs) \citep{Kormann94} with five free parameters: the center position relative to the phase center ($x_L$, $y_L$), the lens strength in the form of the angular Einstein radius ($\theta_E$), the lens shape in the form of the ellipticity of mass distribution ($e_L$), and position angle of the major axis ($\phi_L$). Sometimes we also need to add an external tidal shear with strength $\gamma$ and position angle $\phi_\gamma$. The background sources are represented by one or more S\'ersic \citep{sersic} profiles. 
While simple S\'ersic profiles are unlikely to fully characterise the source structure, they do provide sufficient degrees of freedom to adequately model the data.
These profiles have seven free parameters: the position relative to the lens position ($x_S$, $y_S$), total flux density ($S$), major axis half-light radius ($a_S$), S\'ersic index ($n_S$), axis ratio ($b_S/a_S$), and position angle ($\phi_S$). 

We apply a Markov Chain Monte Carlo (MCMC) fitting procedure using the \texttt{emcee} \citep{emcee} code to sample the parameter space.
In order to decrease the number of visibilities for our lens modeling procedure, we average the data in time in intervals of up to 10 min, unless doing so would cause individual visibilities in the average to have Fourier-plane $uv$ coordinates separated by $>$15 m. This $uv$-plane maximum bin size would cause a source $3''$ from the phase center to suffer a decrease in amplitude by $<2\%$, and thus introduces no serious de-correlation. The limitation on $uv$ averaging only affects the longest baselines that traverse the $uv$ plane the fastest, but the overall noise properties are unaffected because the weights of the averaged visibilities are determined from the sum of the weights of the individual visibilities in each time bin.
For line data we divide the data into 75--250 km s$^{-1}$ velocity bins, depending on the flux distribution and SNR of the line.
For a given lens configuration, we fit the continuum and the CO line channels simultaneously, i.e. fitting multiple source components with one shared set of lens parameters.
We also fit single-channel models of CO with channel widths equivalent to the FWHM of the lines to study the general distribution of the CO gas.
Generally, we run 1000 chains with 1000 burn-in steps and 1000 extra chain steps.  We have verified that the MCMC chains have converged.
We quote the median value of each parameter of the sample as the estimated value and the $1\sigma$ interval as the uncertainty.
When comparing models, we use the Deviance Information Criterion \citep[DIC;][]{Spiegelhalter02} for model selection, which is useful in cases where posterior parameter distributions are determined from MCMC chains.
The best-fit models are selected based on residual images (appendix Figure \ref{fig:residual}), SNRs of channel components, and parameter convergences via the DIC.
The output parameters of the fitting procedures are listed in the appendix Section \ref{sec:paras}.

\section{Analysis}\label{sec:ana}
We find good models for four galaxies, SPT0346-52, SPT0529-54, SPT0532-50 and SPT2147-50, 
where the flux is well recovered and the residuals of the fit are consistent with noise. 
For SPT0532-50, we need two lens components to match the data, and the source position is defined relative to the more massive of the two lenses. For SPT2134-50, we were unable to find an acceptable fit to the data. The structure of the residuals indicates that a simple SIE mass model is not a sufficient representation of the mass distribution in this source. A more complex model of the lens and/or the source is needed for this galaxy, which is beyond the scope of this paper.

\begin{figure}
\plotone{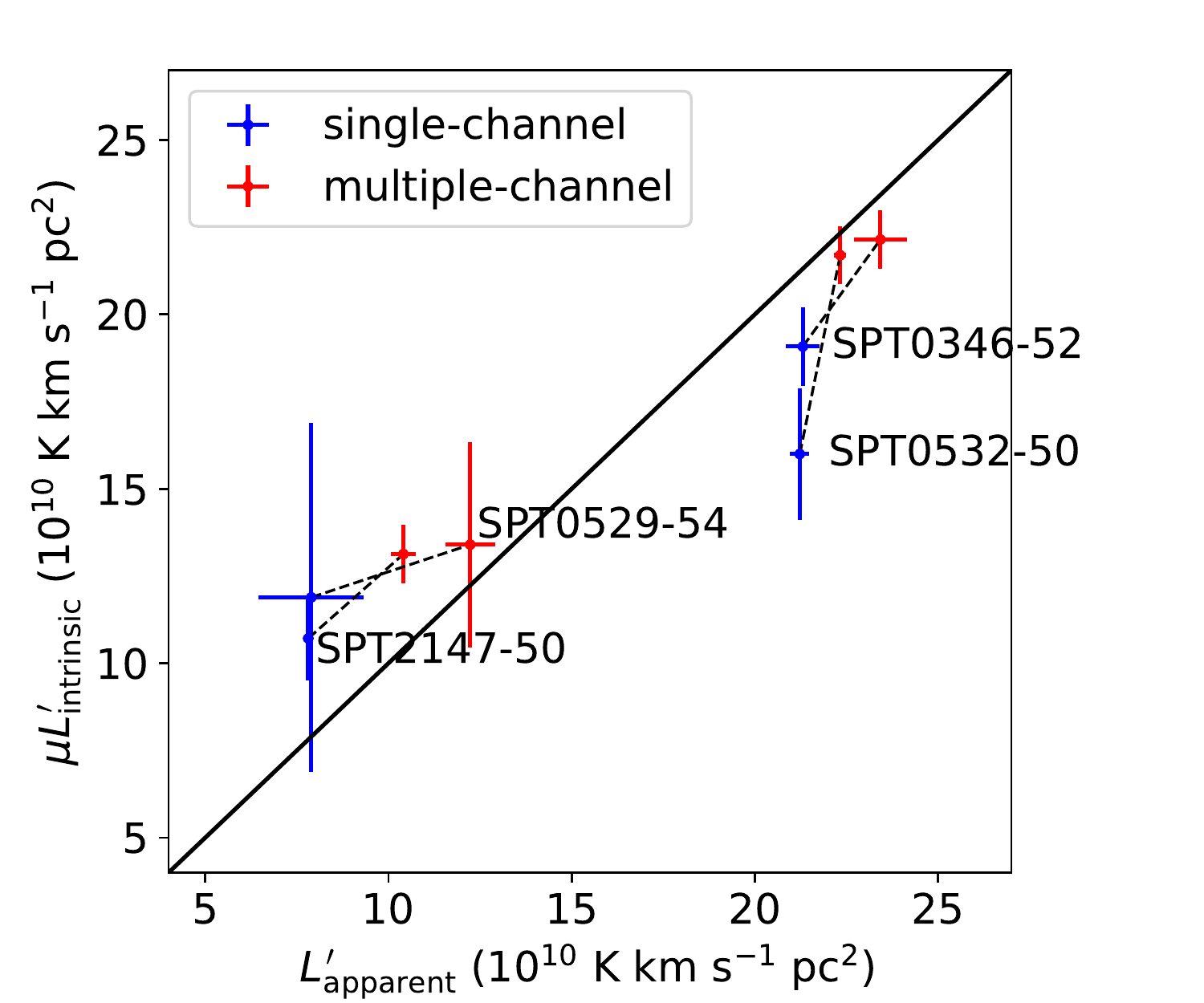}
\caption{The product of magnification and intrinsic CO line luminosity $\mu L^\prime_\mathrm{intrinsic}$ versus the apparent luminosity $L^\prime_\mathrm{apparent}$ for both multiple-channel models and single-channel models.
For multiple-channel models, the $\mu L^\prime_\mathrm{intrinsic}$ values are the sum over all channels of $\mu L^\prime_\mathrm{intrinsic}$ of each individual channel.
The apparent luminosities $L^\prime_\mathrm{apparent}$ are calculated with the same channel widths as those of the corresponding lens models. Points from the same galaxy are connected by a dashed black line.
The apparent line fluxes from the spectra in Figure~\ref{fig:spec} generally agree well with the fluxes derived from our lens modeling. 
\label{fig:mu_comp_L}}
\end{figure}

We quantify the ability of the lens models to recover the line fluxes by comparing the apparent CO line luminosities ($L^\prime_\mathrm{apparent}$) to the product of the magnification and intrinsic line luminosities ($\mu L^\prime_\mathrm{intrinsic}$). We show this comparison in Figure~\ref{fig:mu_comp_L}, finding good agreement  in both the single- and multiple-channel models. $L^\prime$ is the line luminosity expressed in units of K~km~s$^{-1}$~pc$^2$, and is proportional to the line brightness temperature. 
For the multiple-channel model and single-channel model of each source, different band widths are chosen to reach the SNR needed for lens modeling, so the flux coverages are slightly different -- e.g., the faint line wings cannot be included in the multiple-channel models due to low SNR, while this line emission can be accommodated in the single-channel model.
The multiple-channel models better recover the flux at high magnifications, where the impact of differential magnification across the line profile is accounted for. 

\begin{figure*}
\epsscale{1.15}
\plotone{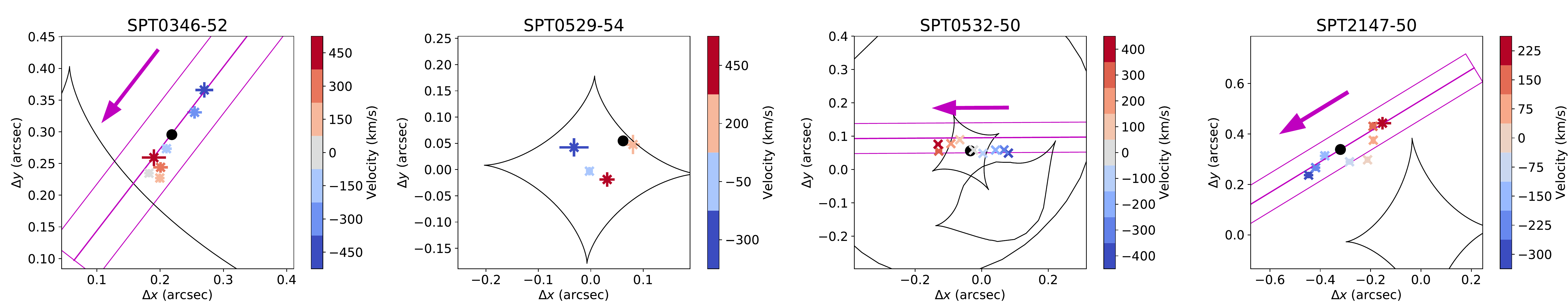}
\plotone{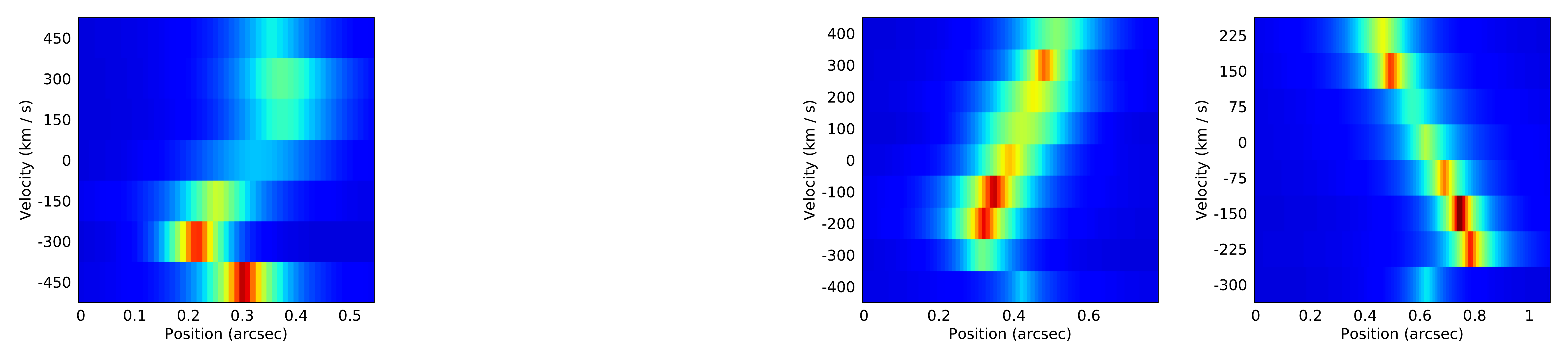}
\plotone{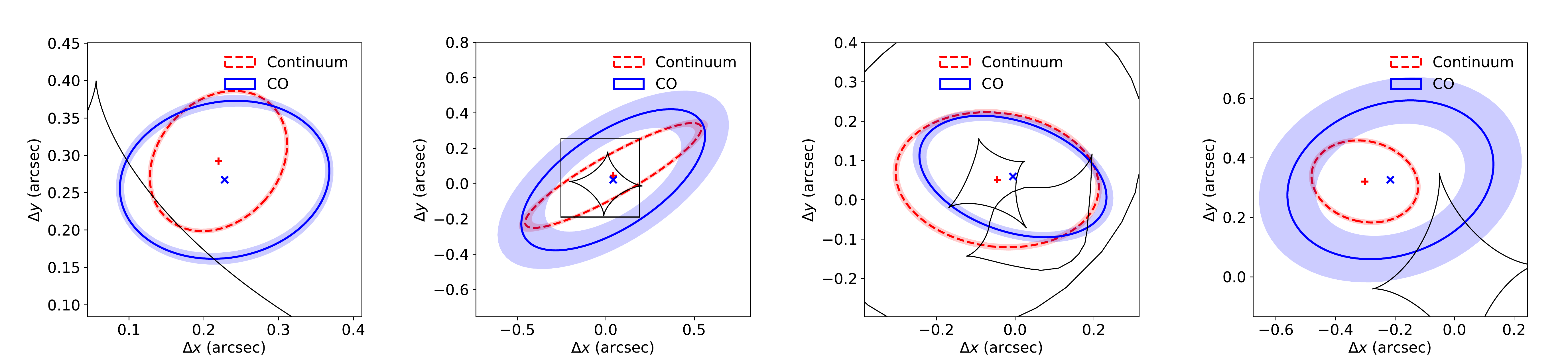}
\caption{
Top row: centroid positions of different CO velocity channels, with 1$\sigma$ uncertainties on the best-fit positions. The black dots represent the dust continuum, and the colored crosses represent the CO lines, with velocities indicated in the colorbar. For SPT0346-52, SPT0532-50 and SPT2147-50, we find clear velocity gradients. For those three galaxies, a magenta slice within which we extract the P-V diagram (middle row) is indicated on the image, and a magenta arrow indicates the positive direction of the position axis.
Middle row: P-V diagrams of SPT0346-52, SPT0532-50 and SPT2147-50 extracted from the slices indicated in the images of the top row. The space for SPT0529-54 is intentionally left blank because no coherent velocity gradient is seen in the reconstruction of this source.
Bottom row: half-light ellipses for CO (blue, solid) and dust continuum (red, dashed) with size uncertainties represented by shaded rings for the intrinsic source. The CO ellipses are derived from single-channel models. 
The observed CO sizes are typically larger than the dust continuum (see also Fig. \ref{fig:linevscont}).
The axes of the bottom row match those of the top row, except for SPT0529-54, where a black box indicating the field-of-view of the top row is shown in the bottom row.
\label{fig:ellips}}
\end{figure*}

The visualizations of our modeling results are shown in Figure \ref{fig:ellips}, where we show (1) the positions of the reconstructed dust emission region and CO velocity channels relative to the lensing caustics, (2) the position-velocity (P-V) diagrams for galaxies where we see a linear position shift of models with different velocity bins -- SPT0346-52, SPT0532-50 and SPT2147-50 -- and (3) the half-light ellipses of the dust and CO emission. 

As seen in the top and middle rows of Figure \ref{fig:ellips}, SPT0346-52, SPT0532-50 and SPT2147-50 show obvious velocity gradients.
SPT0346-52 is likely to be a merger with two sub-components \citep{Spilker15,Litke19}. 
SPT0532-50 and SPT2147-50 show smooth velocity gradients normally typical of galaxy rotation. However, we also cannot rule out that these sources are very close-in galaxy mergers \citep{Hung16}, similar to SPT0346-52. For SPT0529-54, we find no evidence for disk-like rotation or any other obvious velocity gradient despite the elongated structure seen in continuum, integrated CO emission, and individual CO velocity bins. At the depth and resolution of the current data the nature of this source is not yet clear.

We perform our analysis in three stages. First, we investigate the intrinsic sizes derived from lens modeling of the CO and dust observations. 
Second, we compare the magnifications of the dust with the CO, investigate the changes in magnification across the line profiles, and compare magnifications between different CO transitions.
Finally, we perform LVG radiative transfer modeling for both the apparent and the intrinsic line fluxes to study the influence gravitational lensing can have on estimation of physical properties of the lensed galaxies.

\subsection{Source Sizes}\label{sec:size} 
\begin{figure}
\plotone{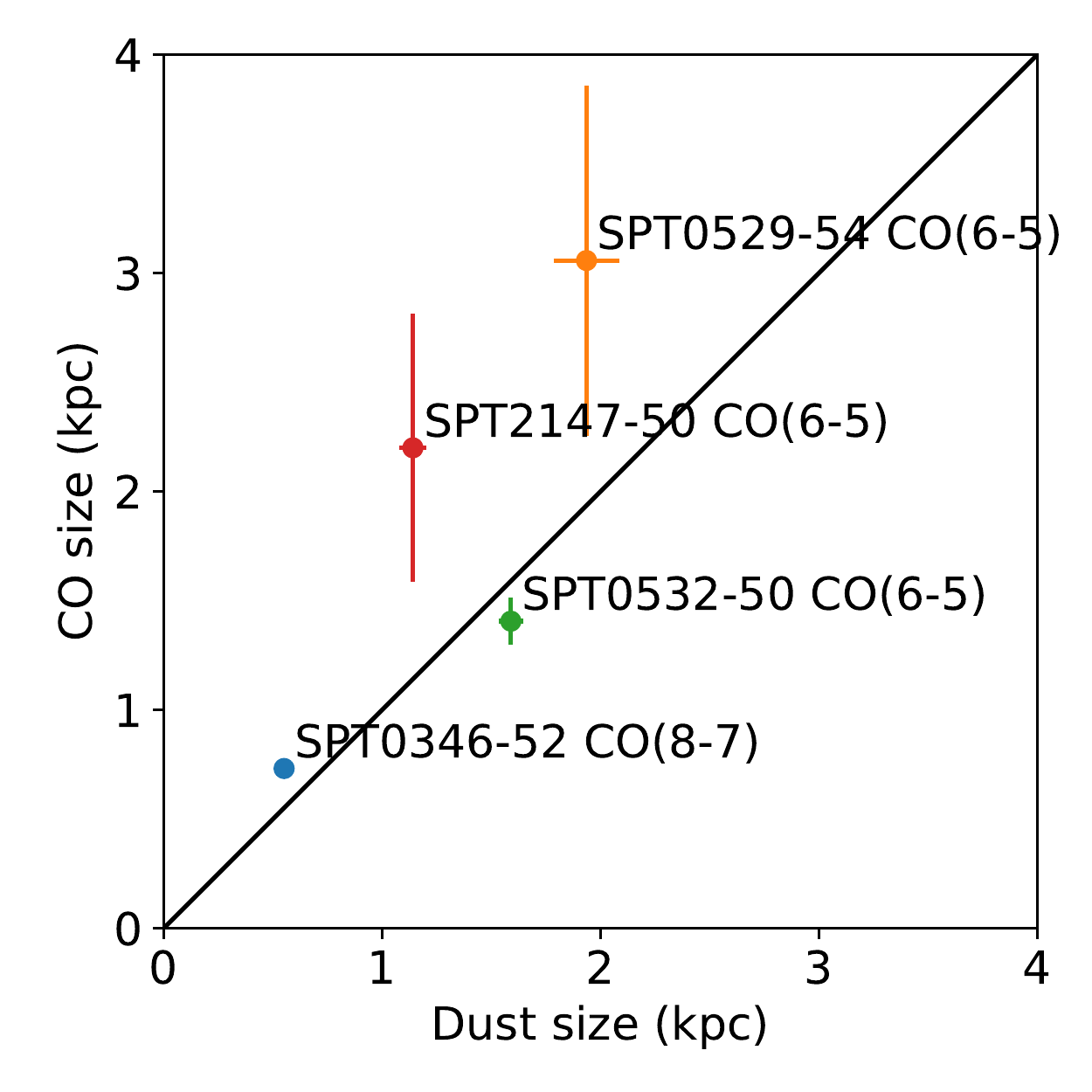}
\caption{
Comparison of circularized sizes of CO line emission with dust continuum. We observe that the CO sizes are larger or comparable than those for the dust continuum.
\label{fig:linevscont}}
\end{figure}

The relative sizes of CO and dust continuum are drawn in the bottom row of Figure \ref{fig:ellips} as half-light ellipses.
To compare the size of CO with dust continuum, we fit the CO line to the lens model in 500--800 km~s$^{-1}$ wide single channels (see Table \ref{tab:linesrcflux} for individual channel widths), which contain the bulk of the total CO emission in all cases. 
We extract the half-light semi-major axis and axis ratio parameters.
We calculate the circularized size and its uncertainty from the MCMC chains as $\sqrt{a_S b_S}$ where $a_S$, $b_S$ are the semi-major and semi-minor axes, respectively. We then compare the circularized size of CO with dust continuum. The comparison is shown in Table \ref{tab:sizecontfluxmass} and plotted in Figure \ref{fig:linevscont}.
We find, for three of the four galaxies, a trend that the size of the CO-emitting region appears larger than the dust continuum, even at high-J transitions of CO(6-5) to CO(8-7). 

If the tendency towards larger CO sizes is confirmed with larger samples, there are several possible explanations for this result.
First, at this wavelength, the dust emission is probably optically thin across the entire source, so the emission is directly proportional to the column density of the dust \citep[e.g.][]{Laursen09}.  Meanwhile, the CO gas is optically thick and the emission arises from all regions where CO is present. Thus, the dust-emitting region can appear smaller than the CO-emitting region in data with limited signal-to-noise. Under standard assumptions about the dust emissivity, gas-to-dust ratio, and the CO-H$_2$ conversion factor, the ALMA dust continuum and ATCA low-J CO data have roughly equal sensitivity to a given molecular gas mass. However, this effect is difficult to test in our data due to our modeling approach -- the faint emission at large radii is not independently constrained because of our parameterized modeling. A modeling approach that affords more freedom in the source plane could offer a more robust test of this scenario.

Second, there may be a temperature gradient in the dust, with cooler temperatures towards the galaxy outskirts away from the regions with the most active star formation. Because the dust emission is also proportional to the effective dust temperature, this effect could also cause lower intensity in the outer regions and thus the smaller apparent size we observe \citep{Galametz12}. While the dust temperature and gas excitation temperature are coupled, the degree to which they are coupled also depends on the gas density, and we note that some observations show that the gas and dust temperatures can be decoupled even at high densities in the presence of strong UV radiation fields and shocks \cite{Koumpia15}. Thus if the effective temperature of the dust emission falls faster than the gas kinetic temperature, this could result in smaller apparent dust than CO sizes.

Finally, it may be true that CO gas has a larger spatial extent than the dust, although this would imply either a spatially varying gas-to-dust ratio (e.g., \citealt{Sandstrom13}) or that the gas and dust are not well-mixed.

\begin{table*}
   \centering
   \caption{
    Derived source properties }
   \begin{tabular}{cccccccccccc}
\toprule
Source & $r_\mathrm{dust}$\tablenotemark{a} & $r_\mathrm{CO}$\tablenotemark{b} & $\nu_\mathrm{cont}$\tablenotemark{c} & $S_\mathrm{cont,aprt}$\tablenotemark{d}  & $S_\mathrm{cont,intr}$\tablenotemark{e}  & $\mu_\mathrm{cont}$\tablenotemark{f}\\
  & (kpc) & (kpc) & (GHz) & (mJy) & (mJy) & \\
\midrule
SPT0346-52 & 0.55 $\pm$ 0.02 & 0.73 $\pm$ 0.04 & 966.7 & 14.75  $\pm$ 0.03 & 2.95  $\pm$ 0.05 & 5.1  $\pm$ 0.1 \\
SPT0529-54 & 1.94 $\pm$ 0.15 & 3.06 $\pm$ 0.80 & 662.0 & 7.59  $\pm$ 0.03 & 0.57  $\pm$ 0.04 & 12.4  $\pm$ 0.6\\
SPT0532-50 & 1.59 $\pm$ 0.06 & 1.41 $\pm$ 0.11 & 661.7 & 9.24  $\pm$ 0.01 & 1.12  $\pm$ 0.04 & 7.9  $\pm$ 0.2  \\
SPT2147-50 & 1.14 $\pm$ 0.06 & 2.20 $\pm$ 0.61  & 659.2 & 4.30  $\pm$ 0.02 & 0.88 $\pm$ 0.05 & 4.8 $\pm$ 0.2 \\
\bottomrule
   \end{tabular}
   \tablecomments{\tablenotemark{a}Dust continuum size. \tablenotemark{b}CO size. \tablenotemark{c}Rest-frame continuum frequency. \tablenotemark{d}Apparent continuum flux density. \tablenotemark{e}Intrinsic continuum flux density. \tablenotemark{f}Continuum magnification.}
   \label{tab:sizecontfluxmass}
\end{table*}

\cite{Spilker15} have already found that low-J CO of DSFGs is more extended than the dust continuum, and note that this effect is also observed in the local universe \citep{Regan01,Leroy08}.
Previous studies indicate the mid-J CO(3-2) emission comes from a region comparable in size to the ongoing star formation both locally \citep{Wilson09} and at high-redshift \citep{Bothwell10,Tadaki17,Chen17,Calistro18}. 
Our study finds that this trend continues at even higher transitions ($J_\mathrm{up}=6$-8) for these (relatively extreme) galaxies. If this trend is real, it may imply that warm and dense gas is more extended than star forming regions and thus that this gas does not have a uniform star formation efficiency.

\subsection{Magnification}
The lensing magnification factor will vary as a function of source position relative to the lens \citep{Hezaveh12}. 
The DSFGs are extended sources, thus different regions of the galaxies can be magnified differently.
Since we find differences in sizes and positions between CO and dust,
as well as among various velocity components of the intrinsic CO, differential magnifications must be present for both situations.

\begin{figure}
\plotone{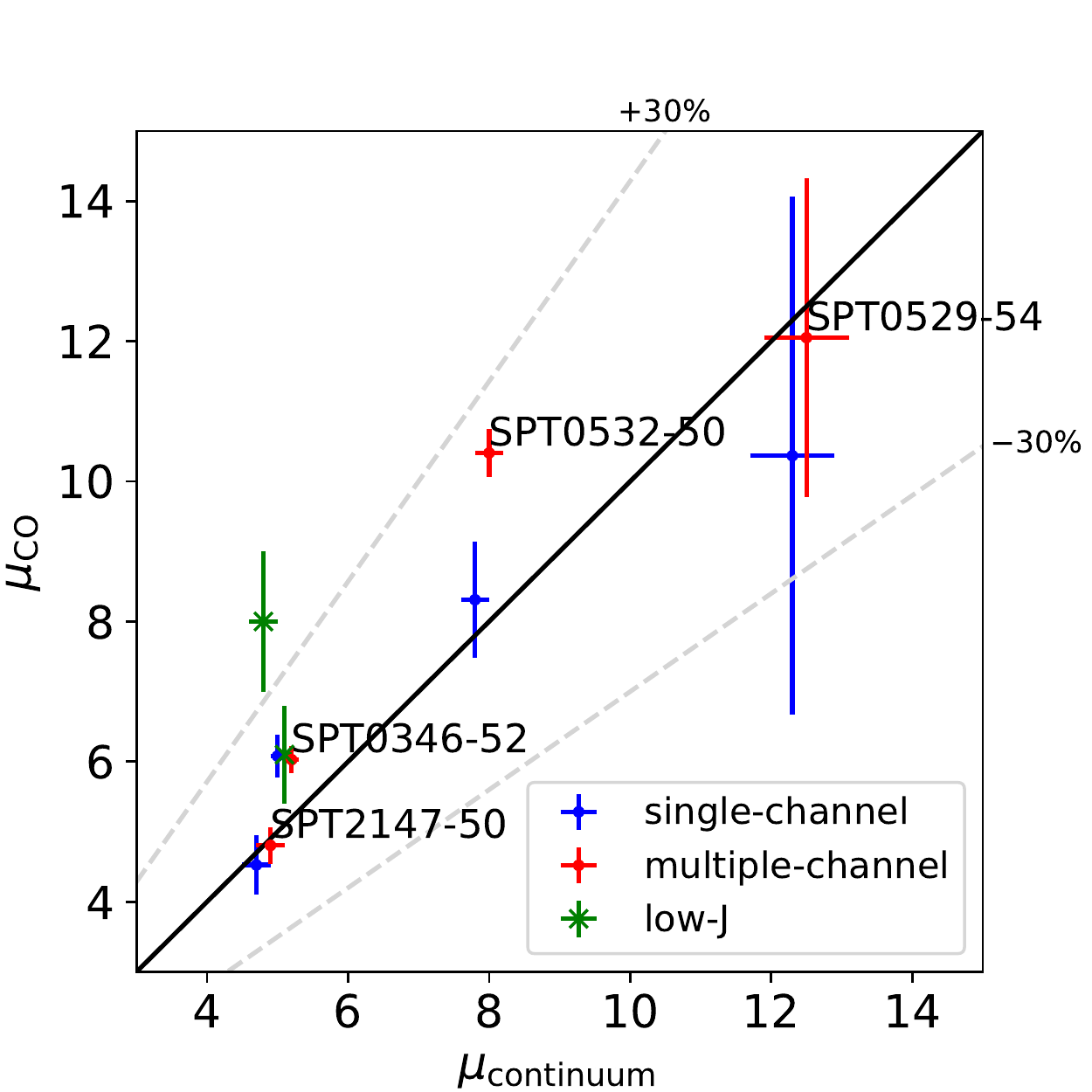}
\caption{Comparison of the magnification factors $\mu_\mathrm{CO}$ versus $\mu_\mathrm{continuum}$ for multiple-channel models (red dots) of our high-J CO, single-channel models (blue dots) of our high-J CO, and low-J CO(2-1) (green crosses, for SPT0346-52 and SPT2147-50). For high-J CO, slight offsets are added to $\mu_\mathrm{continuum}$ to distinguish points from single- and multiple-channel models. All the plotted magnification factors are listed in Table \ref{tab:linesrcflux}. For multiple-channel models, the magnifications are flux-weighted averages of all modeled channels. The two grey dashed lines represent difference of positive and negative 30\% between $\mu_\mathrm{CO}$ and $\mu_\mathrm{continuum}$.
\label{fig:mu_cont_line}}
\end{figure}

\cite{Hezaveh12} simulated a compact and an extended component of a source magnified by the same lens and saw source components with different intrinsic sizes magnified differently. 
\cite{Spilker15} found differences in magnification between dust and molecular gas traced by low-J CO of up to 50\% by analyzing relatively low-resolution observations of SPT0346-52 and SPT0538-50.
In our sample, due to the difference in sizes and positions of CO and dust relative to the caustics (shown in the bottom row of Figure \ref{fig:ellips}), we also see differential magnification between CO and dust for some of our targets, as shown in Figure \ref{fig:mu_cont_line}. 
There is also a deviation of magnifications between single-channel models and multiple-channel models for CO lines (e.g. SPT0532-50 and SPT0529-54) since the multiple models account for more detailed line profiles. This is particularly relevant when the source falls in (or traverses) regions of very high magnification gradients near the lensing caustics.

\begin{figure*}
\epsscale{1.15}
\plotone{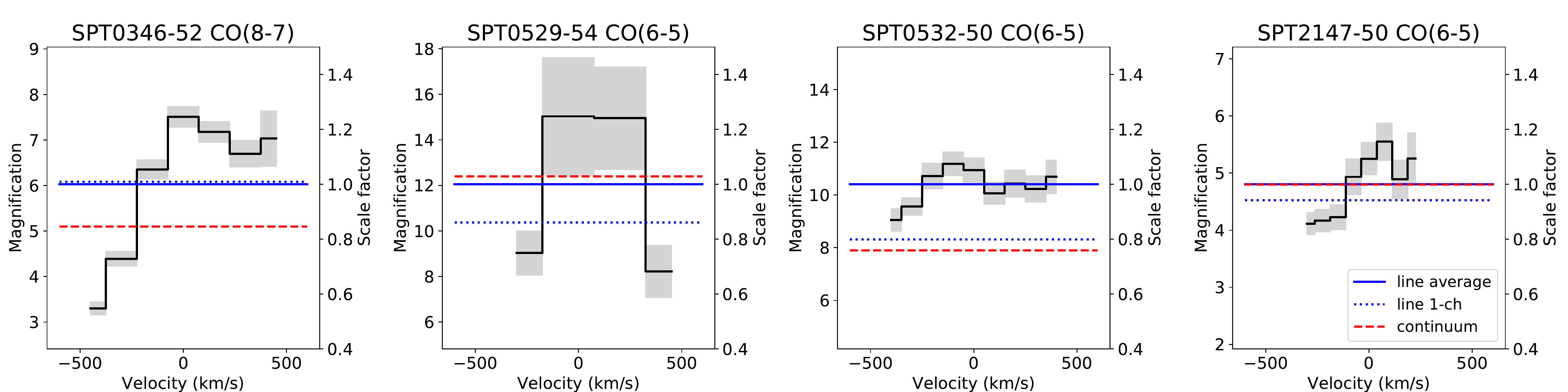}
\caption{Demonstration of differential magnification across the CO line profiles. The black steps represent the magnification of multiple-channel models as a function of velocity across the CO line profiles, with uncertainties represented by the grey shaded areas. The blue solid horizontal lines represent flux-weighted average magnifications from the multiple-channel models.
The blue dotted horizontal lines represent magnifications of single-channel models of the CO line. The red dashed horizontal lines represent magnifications of the dust continuum at rest-frame frequencies near the targeted high-J CO transitions. 
The absolute magnification factors are marked on the left ordinates and the scale relative to the CO multiple-channel averaged magnifications are marked on the right ordinates. 
\label{fig:mu}}
\end{figure*}

Previous studies find that the CO spectral line energy distribution can be affected by differential magnification in low-J CO lines \citep{Deane13,Spilker15,Rivera18}. Because we model separate channels through velocity space, we also find differential magnification across our high-J CO line profiles, as shown in Figure \ref{fig:mu}. 
We see up to factor of 2 magnification variations across the line profile (e.g. in SPT0346-52), but in general it is no more than 30\%.
The flux-weighted average magnifications are indicated in blue solid horizontal lines in Figure \ref{fig:mu}, which may vary from the magnifications derived from single-channel models (blue dotted horizontal lines) and dust continuum (red dashed horizontal lines). 

\begin{figure*}
\epsscale{1.15}
\plotone{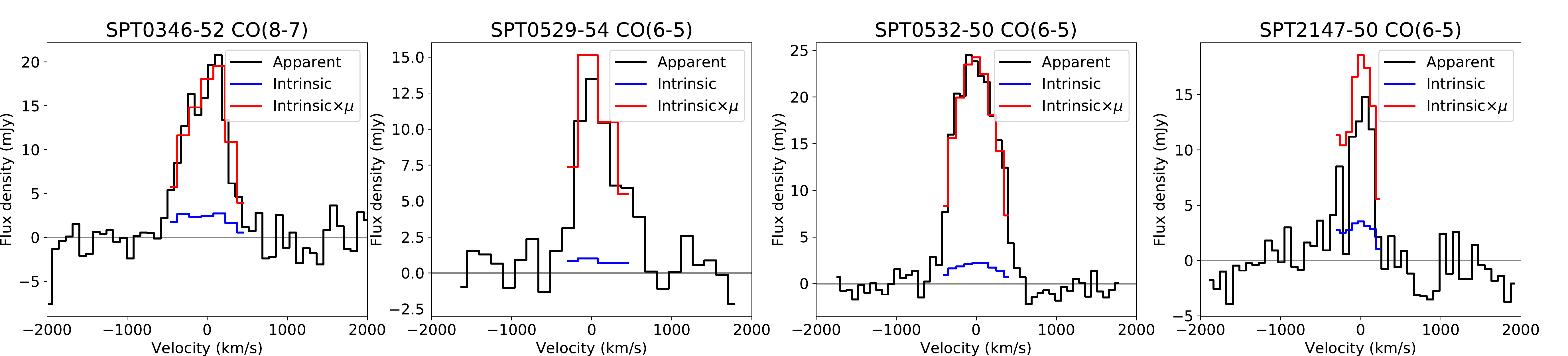}
\caption{Comparison of the apparent spectra (i.e., uncorrected for lensing magnification) versus the intrinsic spectra (i.e., accounting for the magnification). The black apparent spectra are the same as in Figure \ref{fig:spec}. The blue spectra are the lensing-corrected spectra, and the red spectra show the product of the magnifications of each channel and the intrinsic fluxes. The channelization between apparent and intrinsic spectra are not identical because we require wider channels in order to have sufficient SNR to perform lens modeling, limiting the velocity resolution of the intrinsic spectra.
\label{fig:specimgsrc}}
\end{figure*}

To test the consistency of the apparent spectrum and the reconstructed line fluxes from the lens models, we plot the intrinsic line fluxes multiplied by the derived magnifications in Figure \ref{fig:specimgsrc}. The apparent line fluxes are well recovered across the line profiles in our models.

We list the derived magnifications and source fluxes in Table \ref{tab:sizecontfluxmass} for the dust continuum and Table \ref{tab:linesrcflux} for the CO line.  For the line fluxes, we list line fluxes for both single-channel and multiple-channel models. 
For multiple-channel models, the magnifications are the flux-weighted average of all channels. 
We also list the line luminosities in units of K~km~s$^{-1}$~pc$^2$. 

For SPT0346-52 and SPT2147-50, we also compare our lens models of high-J CO emission to similar lens models of high-resolution CO(2-1) data observed with ATCA. For SPT0346-52, these data were presented in \citet{Spilker15}; here, we update that lensing model with the best-fit lens parameters from the high-J ALMA data.  The line fluxes and magnification factors of the two models are listed in Table \ref{tab:linesrcflux} and shown in Figure \ref{fig:mu_cont_line}. We again see differential magnification between CO transitions in these two galaxies, especially in SPT2147-50. For SPT0346-52 the magnifications of the CO(2-1) and CO(8-7) transitions are similar (to within $\sim$2\%), while for SPT2147-50 we see CO(6-5) magnified by a factor $\sim$5 and CO(2-1) magnified by $\sim$8$\times$, $\sim$60\% higher than the high-J CO magnification.
 We analyze the effects of this feature further in Section \ref{sec:lvg} using LVG radiative transfer modeling.

\begin{table*}
   \centering
   \caption{Intrinsic line fluxes and magnifications}
   \begin{tabular}{cccccccc}
\toprule
Source & Line & $N_\mathrm{ch}$\tablenotemark{a} & Width\tablenotemark{b} & $S\mathrm{d}v$ & $L^\prime_\mathrm{CO}$ & $\mu$ \\
  &   &   & (km s$^{-1}$) & (Jy km s$^{-1}$) & ($10^{10}$ K km s$^{-1}$ pc$^2$) &   &   \\
\midrule
 SPT0346-52 & CO(2-1) & 5 & 150 & 0.26  $\pm$ 0.02 & 7.20  $\pm$ 0.53 & 6.1  $\pm$ 0.7 \\
 & CO(8-7) & 7 & 150 & 2.10  $\pm$ 0.04 & 3.67  $\pm$ 0.07 & 6.0  $\pm$ 0.2 \\
 &  & 1 & 800 & 1.80  $\pm$ 0.06 & 3.14  $\pm$ 0.10 & 6.1  $\pm$ 0.3 \\
SPT0529-54 & CO(6-5) & 4 & 250 & 0.80  $\pm$ 0.09 & 1.11  $\pm$ 0.13 & 12.1  $\pm$ 2.3 \\
 &  & 1 & 700 & 0.82  $\pm$ 0.18 & 1.15  $\pm$ 0.25 & 10.4  $\pm$ 3.7 \\
SPT0532-50 & CO(6-5) & 9 & 100 & 1.48  $\pm$ 0.03 & 2.08  $\pm$ 0.04 & 10.4  $\pm$ 0.3 \\
 &  & 1 & 800 & 1.36  $\pm$ 0.08 & 1.93  $\pm$ 0.12 & 8.3  $\pm$ 0.8 \\
SPT2147-50 & CO(2-1) & 3 & 150 & 0.20  $\pm$ 0.01 & 2.95  $\pm$ 0.23 & 8.0  $\pm$ 1.0 \\
 & CO(6-5) & 8 & 75 & 1.65  $\pm$ 0.06 & 2.73  $\pm$ 0.10 & 4.8  $\pm$ 0.3 \\
 &  & 3 & 150 & 1.43  $\pm$ 0.09 & 2.37  $\pm$ 0.14 & 4.5  $\pm$ 0.4 \\
 &  & 1 & 500 & 0.82  $\pm$ 0.14 & 1.36  $\pm$ 0.24 & 5.2  $\pm$ 1.4 \\
\bottomrule
   \end{tabular}
\tablecomments{
\tablenotemark{a}Number of channels. \tablenotemark{b}Channel width.}
    \label{tab:linesrcflux}
\end{table*}

\subsection{Radiative transfer modeling}\label{sec:lvg}

One commonly-used application of well-sampled CO ladders is to do large velocity gradient (LVG) radiative transfer modeling in order to constrain the temperature and density of the CO-emitting gas. To test how different line ratios between the apparent and intrinsic images affect the study of physical properties of the gas, we conduct LVG modeling using both apparent and intrinsic line ratios to evaluate the differences between them. 
Here, we do a simple study with two of our sources where we have spatially resolved multiple CO lines (SPT0346-52 and SPT2147-50). SPT2147-50 shows significant differential magnification between the spatially resolved CO transitions, 
while SPT0346-52 does not. We compare this to the case where one has observed many lines at low spatial resolution, and so must assume the same magnification between all of them.
For the CO(8-7) line of SPT0346-52, we checked the flux with another ALMA observation of this same transition (2013.1.00722.S, PI: M.~Aravena) and found a mismatch in the sense that our line fluxes are higher, likely due to the use of different flux calibrators, so we added an additional 25\% uncertainty to the flux of that line for this comparison.

\begin{figure}
\epsscale{1.15}
\plotone{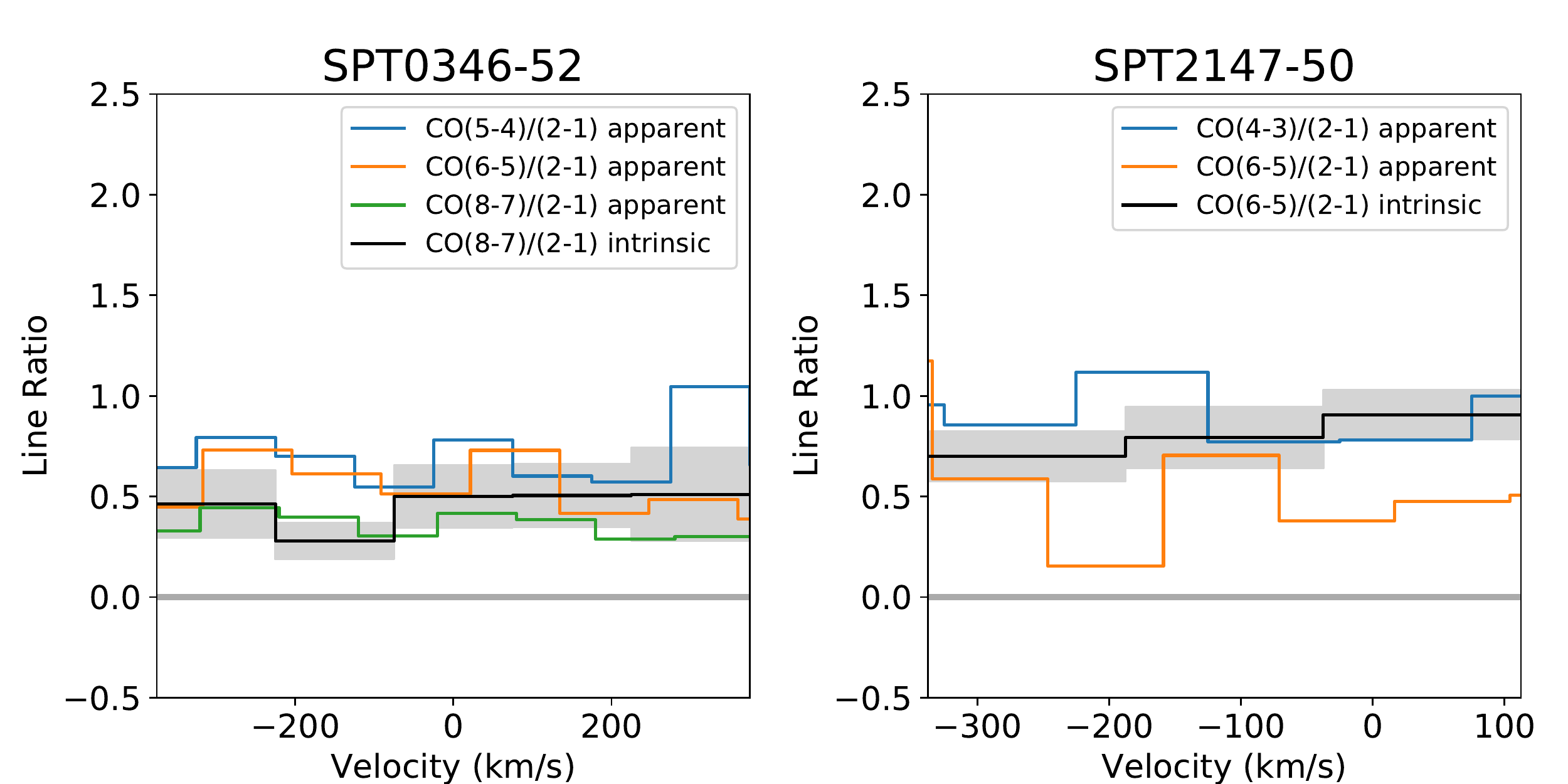}
\caption{$L^\prime_\mathrm{CO}$ line ratios across the line profiles for SPT0346-52 and SPT2147-50. The black lines are the intrinsic line ratios from the lens models,  with uncertainties represented by grey shaded areas. The colored lines represent the apparent line ratios. To make sure the intrinsic ratio profiles are calculated within the same velocity bins despite the large difference in data signal-to-noise, we made 150~km~s$^{-1}$ channel models for both sources (see Table \ref{tab:linesrcflux}). Although some differential magnification is seen across the line profiles (Figure~\ref{fig:mu}), the line ratios are fairly constant as functions of velocity. For SPT2147-50, the overall differential magnification between CO(6-5) and CO(2-1) is apparent in the global offset of the line ratio between the apparent and intrinsic images.
\label{fig:ratio}}
\end{figure}

First, we investigate whether the ratios of $L^\prime_\mathrm{CO}$ line luminosities between different transitions vary across the line profiles due to differential magnification. Because of the large differences in signal-to-noise between the low- and high-J data, we cannot exactly match the channelization of the data for both lines. For SPT0346-52, we re-fit the CO(2-1) data from \citet{Spilker15}, matching the 150 km s$^{-1}$ channels used in the CO(8-7) modeling; the outermost channels in the line wings are too faint to model. For SPT2147-50, we re-fit the CO(6-5) data in 150 km s$^{-1}$ channels, matching the resolution achievable for the CO(2-1) data. In Figure \ref{fig:ratio} we plot both the intrinsic and apparent line ratios of each higher transition relative to CO(2-1) for the two galaxies. We find the line ratios remain reasonably constant across the line profiles, indicating lensing distortion of the line profiles does not significantly affect the estimation of physical conditions using line luminosities. 

While the apparent and intrinsic line ratios are similar for SPT0346-52, they differ by $\sim$60\% for SPT2147-50. This difference is directly due to the large difference in magnification between CO(6-5) and CO(2-1) in this galaxy.
Because the CO(2-1) magnification is higher overall than CO(6-5), the true CO line ratio in SPT2147-50 is higher than would be inferred from the apparent line ratio, which influences the conclusions drawn about the gas physical conditions in the absence of spatially-resolved data.

To further investigate this discrepancy in line ratios, we perform simple radiative transfer modeling under the Large Velocity Gradient (LVG) approximation \citep{Sobolev60,Castor70,Scoville74,Goldreich74}. CO lines often have substantial optical depths due to the high abundance of CO. For optically thick emission, radiative trapping is important, requiring an iterative calculation of the CO level populations and the emergent line emission. Here, we use the \texttt{RADEX} radiative transfer code \citep{cradex}, combined with an MCMC algorithm to sample the highly degenerate parameter space. A similar approach can be found in \cite{Yang17}.  We run the MCMC chains with three free variables, (1) $N_\mathrm{CO}/\Delta v$, the column density of CO divided per velocity interval, with a flat prior in the range $10^{14}$--$10^{19}$ cm$^{-2}$ (km s$^{-1}$)$^{-1}$; (2) $T_\mathrm{kin}$, the kinetic temperature of the gas, with a flat prior in the range $T_\mathrm{CMB}$--$300$ K; and (3) $n_\mathrm{H_2}$, the number density of molecular hydrogen, with flat prior in the range $10$--$10^6$ cm$^{-3}$. 
Following the \texttt{RADEX} documentation, we apply a prior that the line optical depth $\tau$ be less than 100, and set the background radiation temperature to the cosmic microwave background (CMB) temperature $T_\mathrm{CMB}=2.73(1+z)$ K. 

Starting from an initial set of parameters, we calculate the emergent CO line fluxes, and hence compute the line ratios of each higher transition over the lowest transition available using data in Table \ref{tab:lineimgflux}, e.g. for SPT0346-52, we calculate CO(8-7)/CO(2-1), CO(6-5)/CO(2-1), and CO(5-4)/CO(2-1). With the corresponding line ratios from the observations, we can compute the likelihood that the given set of parameters reproduce the data.

We model under three scenarios. 
First, we construct a model using all available lines\footnote{CO(8-7), CO(6-5), CO(5-4) and CO(2-1) for SPT0346-52; CO(6-5), CO(4-3), CO(2-1) for SPT2147-50}, assuming all lines are magnified equally such that the apparent and intrinsic line ratios are equal.
Second, we model the intrinsic line ratios (corrected for differential magnification) using the two available lines for each galaxy\footnote{CO(8-7) and CO(2-1) for SPT0346-52; CO(6-5) and CO(2-1) for SPT2147-50}.
Finally, we model the apparent line ratios for this same pair of transitions for each source.
By comparing the first and the third scenarios, we study how the derived physical constraints change depending upon the number of lines available, ignoring the effects of differential magnification.
By comparing the second and the third scenarios, we study how the derived physical constraints may be biased by differential magnification, by modeling the same apparent and intrinsic line ratios. 

\begin{figure*}
\plotone{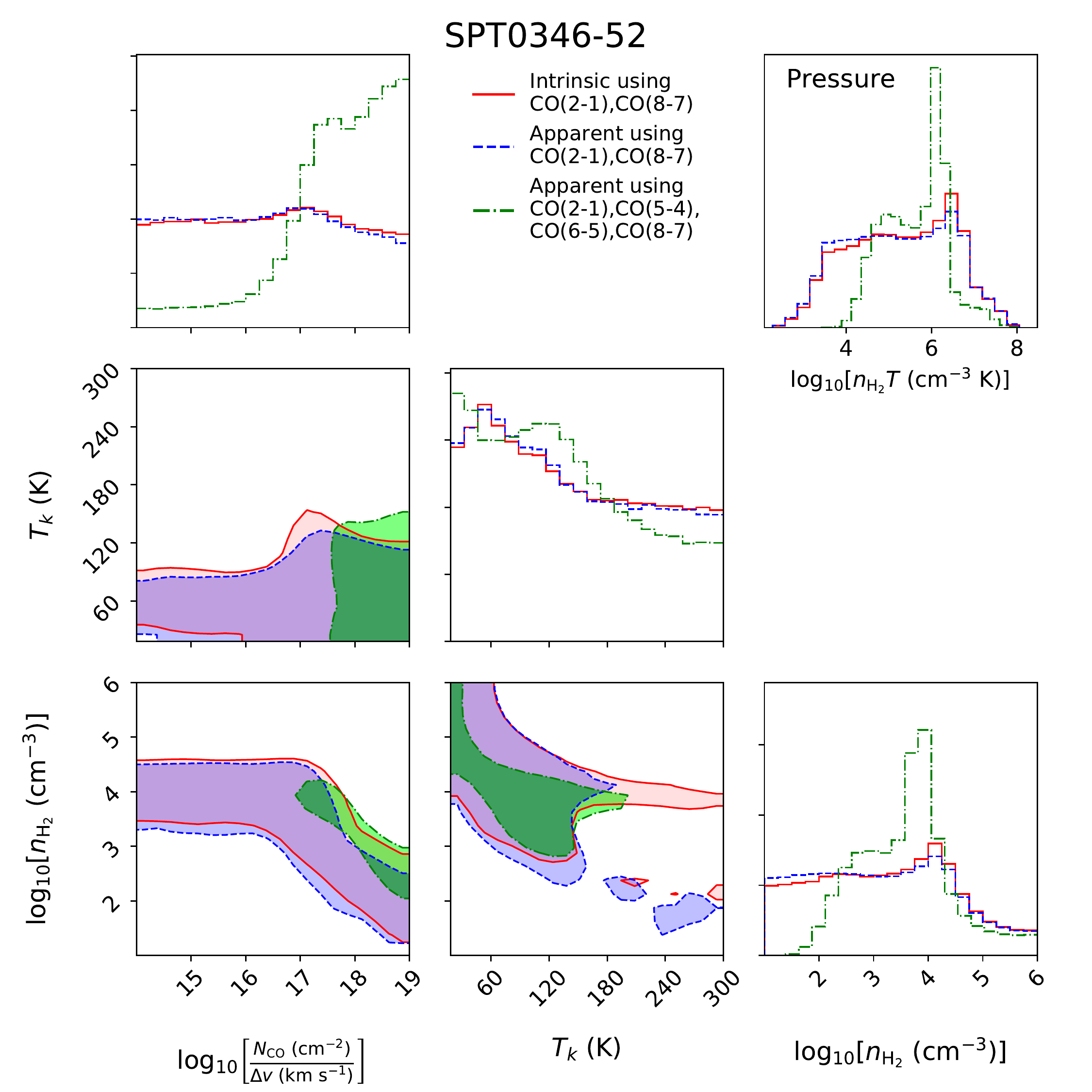}
\caption{LVG modeling results for SPT0346-52. We show the two-dimensional parameter covariances between column density, temperature, and number density, with the marginalized one-dimensional distributions along the diagonal. We also include a histogram of the pressure $p/k=n_\mathrm{H_2}T$. 
The red solid lines and shaded regions show the parameter distributions using the two intrinsic line fluxes available.
The blue dashed lines and shaded regions show the results using the same lines but the apparent (not corrected for differential magnification) line ratios.
Finally, the green dash-dotted lines and shaded regions show the results using the apparent line ratios and all available lines (also not corrected for differential magnification).
For the two-dimensional covariance plots, contours enclose 68\% (1$\sigma$) of the total likelihood.
This figure shows that, for SPT0346-52, differential magnification does not significantly affect the inferred physical conditions (the red and blue regions are quite similar), but that more lines observed lead to better constrained parameter distributions (blue vs green regions).
\label{fig:radex1}}
\end{figure*}

\begin{figure*}
\plotone{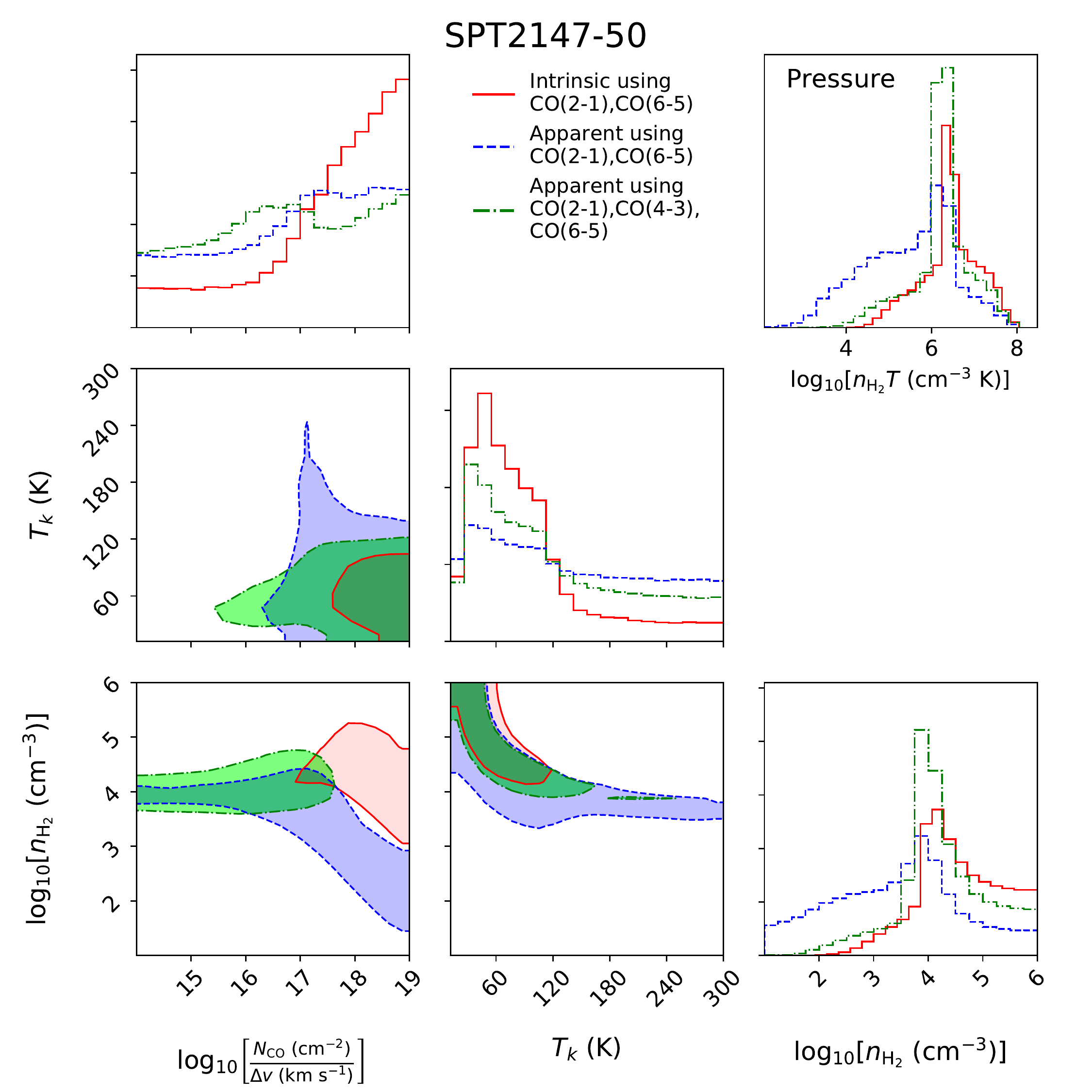}
\caption{Same as Fig. \ref{fig:radex1}, but for SPT2147-50. In this case, differential magnification does affect the inferred parameter distributions (difference between blue and red regions). For this source, a higher gas density is obtained when differential magnification is accounted for.
\label{fig:radex5}}
\end{figure*}

The parameter distributions of the MCMC chains are shown in Figures \ref{fig:radex1} and \ref{fig:radex5}, generated by the code from \cite{corner}. We note that the unbounded parameter distributions seen in these Figures are generic features of LVG modeling \citep{Spilker14}; increased parameter ranges would not lead to bounded distributions.
From the LVG modeling, we find SPT0346-52 and SPT2147-50 yield disparate results. 

SPT0346-52 is found to yield similar physical properties if studied with the same lines for either the apparent or intrinsic ratios, as expected from the negligible differential magnification. Modeling with additional CO(5-4) and CO(6-5) lines returns more constrained parameters, and is consistent with the studies with fewer lines. As the magnifications of the different CO lines are quite similar, the number of lines considered dominates the LVG modeling process. For this galaxy, acquiring spatially unresolved data for as many lines as possible is the preferred method for constraining the gas physical conditions.

SPT2147-50, on the other hand, shows the opposite result -- namely, that the inferred physical properties do depend on whether the analysis is performed using apparent or intrinsic line ratios. As with SPT0346-52, the apparent analysis using all available lines yields results consistent with, but more tightly constrained than, the modeling using fewer lines. However, the results using the intrinsic line ratios do show differences compared to the apparent analysis. For this particular galaxy, the true gas density is higher than would otherwise be inferred, a direct consequence of the higher overall magnification of CO(2-1) than CO(6-5). For this galaxy, a proper understanding of the gas physical conditions requires spatially resolving as many lines as possible in order to account for the substantial differential magnification between transitions.

Therefore, both the number of available line luminosities with different CO transitions and the differential magnification between different transitions can affect the physical condition estimates of the gas. Unfortunately, there is no way to know a priori whether observing a large number of lines at low spatial resolution or fewer lines at higher resolution is the more promising strategy.
Previous LVG studies of high-z DSFGs \citep[e.g.][]{Spilker14,Yang17} suggest a low-excitation component and a high-excitation component in the systems, but few of them also perform detailed lens modeling of the observed lines. There may be no single optimal tradeoff between the number of lines observed and the precision of the lens model, but in general exploring a wider range of CO transitions and performing detailed lens modeling for each CO transition will be important for future analyses.
However, we expect that obtaining high spatial resolution imaging in order to model the lensing magnification is obviously important when considering tracers that arise from different physical conditions (e.g., low-J CO and high-J CO), while it should be less critical for more closely related tracers (e.g., adjacent CO transitions).
We suggest that observing two CO lines with widely separated $J_\mathrm{up}$ is an efficient method to determine the relative importance of differential magnification for the interpretation of CO line ratios.

\section{Conclusions}\label{sec:con}
We perform parameterized, visibility-based gravitational lens modeling 
on five DSFGs discovered by the SPT survey. 
We use ALMA observations of high-J CO line emission ($J_\mathrm{up}=6$, 7, 8) and dust continuum  to study the relative sizes and magnifications of molecular gas and dust.
We find the high-J CO spectra have similar shapes to lower-J CO spectra.
Our modeling accurately reproduces the observations of four of the five lensed galaxies. We find:

(1) The physical extent of the region traced by the high-J CO lines tends to be comparable to or larger in size than the dust continuum at rest frequencies near the CO transitions. 
This is different from ``normal'' local and high-z galaxies where mid-J CO generally has an extent similar to the dust \citep[e.g.][]{Wilson09,Spilker15,Tadaki17,Chen17,Calistro18}.
Since high-J CO lines trace warmer and denser gas than lower-J transitions, their large size indicates that star-forming conditions are widespread within these galaxies.

(2) We find differential magnification between dust continuum and CO molecular gas, across the CO line profiles, and between different CO transitions. This differential magnification is due to positional offsets and size differences between different components of the galaxies. The magnitude of the differential magnification reaches as high as a factor of 2, but is generally no more than $\sim$30\%.

(3) We find velocity gradients in three of our sources. 
Two of them (SPT0532-50 and SPT2147-50) show gradients consistent with ordered rotation. 
The third one, SPT0346-52, is likely to be a merger, based in part on other data analyzed by \cite{Litke19}.

(4) For SPT0346-52 and SPT2147-50, we compare the intrinsic and apparent line ratios between CO(2-1) and higher-J CO and quantify the effect of differential magnification. For SPT0346-52, differential magnification is negligible, while for SPT2147-50, the CO(2-1) emission is magnified $\sim$60\% more highly than CO(6-5), indicating a higher intrinsic CO excitation than the apparent line ratio would indicate.

(5) We perform LVG modeling to study the physical properties of two galaxies (SPT0346-52 and SPT2147-50) under three scenarios: (a) Using the apparent line ratios using all available lines (4 lines for SPT0346-52 and 3 lines for SPT2147-50); (b) using the intrinsic line ratios using the two lines in each source for which we have lens models; (c) using these same two lines but the apparent line ratios for each source.

For SPT0346-52, scenario (a) gives a more constrained result than scenarios (b) and (c), while scenarios (b) and (c) give similar results. Since the intrinsic high-J/low-J line ratio is comparable to the apparent line ratio in SPT0346-52, this indicates the number of lines input into LVG modeling has a greater impact than differential magnification on determination of physical parameters.

For SPT2147-50, scenario (b) gives a divergent result from scenarios (a) and (c), while scenarios (a) and (c) give similar results. Since the intrinsic high-J/low-J line ratio is higher than the apparent line ratio in SPT2147-50, this indicates that if the difference in magnification is too great between lines, its effect can overwhelm the bias from the incompleteness of the CO ladder observations.

These cases are two examples that together illustrate potential challenges for the interpretation of CO lines observations in lensed DSFGs. 
We suggest that observing two CO lines with widely separated $J_\mathrm{up}$ is an efficient method to determine the relative importance of differential magnification for the interpretation of CO line ratios.
Future analyses including more lines with lens models will be necessary to better quantify what data are required to derive robust constraints.

\acknowledgments
C.D. thanks the University of Florida for support through a Graduate School Fellowship.
J.S.S. thanks the McDonald Observatory at the University of Texas at Austin for support through a Smith Fellowship.
This paper makes use of the following ALMA data: ADS/JAO.ALMA \#2011.0.00957.S, \#2013.1.00880.S and \#2015.1.00117.S.
ALMA is a partnership of ESO (representing its member states), NSF (USA) and NINS (Japan), together with NRC (Canada), MOST and ASIAA (Taiwan), and KASI (Republic of Korea), in cooperation with the Republic of Chile. The Joint ALMA Observatory is operated by ESO, AUI/NRAO and NAOJ.
The National Radio Astronomy Observatory is a facility of the National Science Foundation operated under cooperative agreement by Associated Universities, Inc.
The Australia Telescope Compact Array is part of the Australia Telescope National Facility which is funded by the Australian Government for operation as a National Facility managed by CSIRO.
The authors acknowledge University of Florida Research Computing for providing computational resources and support that have contributed to the research results reported in this publication. 
The Flatiron Institute is supported by the Simons Foundation.

\appendix
\section{Lens model residual maps}
\begin{figure*}
\epsscale{1.15}
\plotone{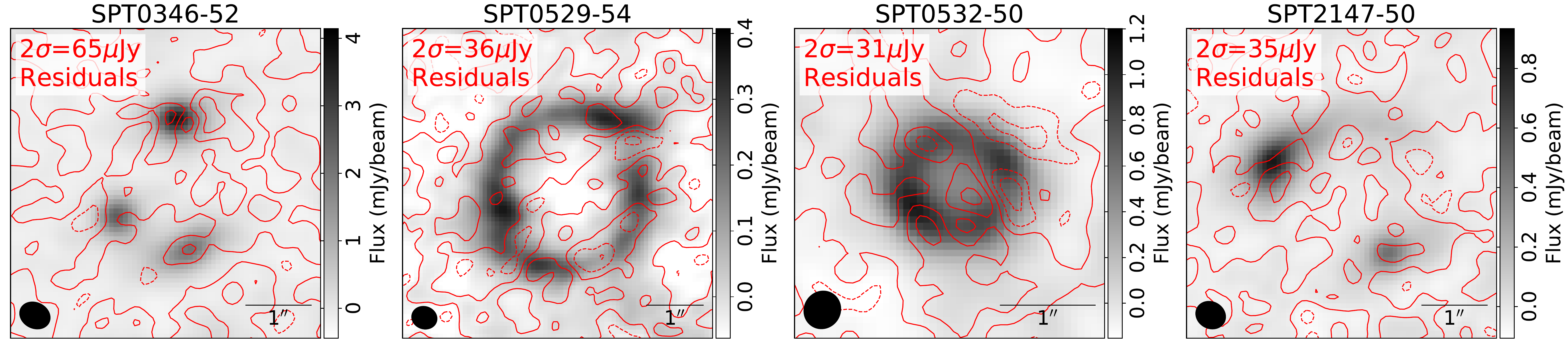}
\caption{The continuum data (greyscale) and data$-$model residual maps (red contours) for our best-fit models. The residual contours start at 0 and have steps of  $\pm2\sigma$, with 2$\sigma$ flux values indicated in the top left corners of each panel. Zero and positive contours are in solid lines and negative contours are in dashed lines. The synthesized beams are indicated in the lower left corners and $1''$ scale bars are indicated in the lower right corners. 
\label{fig:residual}}
\end{figure*}

We here present the residual maps of our best-fit continuum models (Figure \ref{fig:residual}) to illustrate the quality of our models.

\section{Lens model parameters}\label{sec:paras}
The output parameters for the lenses, continuum sources and CO sources are listed in Tables \ref{tab:lenspara} to \ref{tab:sourceparaSPT2147dv150}. See Section \ref{sec:len} for parameter descriptions.

\begin{table*}
\scriptsize
   \centering
   \caption{Lens parameters}
   \begin{tabular}{@{} lcccccccc @{}}
\toprule
Galaxy & $x_L$\tablenotemark{a} & $y_L$\tablenotemark{b} & $\theta_{E,L}$\tablenotemark{c} & $e_L$\tablenotemark{d} & $\phi_L$\tablenotemark{e} & $\gamma$\tablenotemark{f} & $\phi_\gamma$\tablenotemark{g} \\
--- & (arcsec) & (arcsec) & (arcsec) & --- & (deg CCW from E) & --- & (deg CCW from E) \\
\midrule
SPT0346-52 & -0.568 $\pm$ 0.006 & -0.369 $\pm$ 0.005 & 1.00 $\pm$ 0.10 & 0.58 $\pm$ 0.01 & 71.8 $\pm$ 0.5 & 0.114 $\pm$ 0.003 & 130 $\pm$ 2 \\
SPT0529-54 & -0.078 $\pm$ 0.006 & -0.154 $\pm$ 0.006 & 1.32 $\pm$ 0.12 & 0.20 $\pm$ 0.01 & 87.7 $\pm$ 1.3 & --- & --- \\
SPT0532-50 & -0.066 $\pm$ 0.003 & -0.274 $\pm$ 0.003 & 0.43 $\pm$ 0.06 & 0.36 $\pm$ 0.02 & 47.1 $\pm$ 1.5 & 0.070 $\pm$ 0.004 & -17 $\pm$ 2 \\
           & -0.235 $\pm$ 0.007 & -0.041 $\pm$ 0.006 & 0.12 $\pm$ 0.03 & 0.15 $\pm$ 0.02 & 154.8 $\pm$ 6.7 & --- & --- \\
SPT2147-50 & 0.791 $\pm$ 0.011 & -0.708 $\pm$ 0.012 & 1.20 $\pm$ 0.11 & 0.35 $\pm$ 0.01 & 5.3 $\pm$ 0.9 & --- & --- \\
\bottomrule
   \end{tabular}
 \tablecomments{\tablenotemark{a}\tablenotemark{b}Position relative to the phase center. \tablenotemark{c}Einstein radius. \tablenotemark{d}Lens ellipticity. \tablenotemark{e}Position angle of the major axis. \tablenotemark{f}External shear. \tablenotemark{g}Shear angle.}
   \label{tab:lenspara}
\end{table*}

\begin{table*}
\scriptsize
   \centering
   \caption{Continuum Source parameters}
   \begin{tabular}{@{} lcccccccc @{}}
\toprule
Galaxy & $x_S$\tablenotemark{a} & $y_S$\tablenotemark{b} & $S_\mathrm{cont}$\tablenotemark{c} & $a_S$\tablenotemark{d} & $n_S$\tablenotemark{e} & $b_S/a_S$\tablenotemark{f} & $\phi_S$\tablenotemark{g} & $\mu$\tablenotemark{h} \\
 & (arcsec) & (arcsec) & (mJy) & (arcsec) &  &  & (deg CCW from E) &  \\
\midrule
SPT0346-52 & 0.218 $\pm$ 0.003 & 0.295 $\pm$ 0.006 & 2.95 $\pm$ 0.05 & 0.106 $\pm$ 0.003 & 1.48 $\pm$ 0.08 & 0.77 $\pm$ 0.02 & 51.5 $\pm$ 3.2 & 5.1 $\pm$ 0.1 \\
SPT0529-54 & 0.062 $\pm$ 0.009 & 0.055 $\pm$ 0.006 & 0.57 $\pm$ 0.04 & 0.533 $\pm$ 0.039 & 0.81 $\pm$ 0.10 & 0.20 $\pm$ 0.01 & 29.5 $\pm$ 0.7 & 12.4 $\pm$ 0.6 \\
SPT0532-50 & -0.034 $\pm$ 0.002 & 0.056 $\pm$ 0.002 & 1.12 $\pm$ 0.04 & 0.274 $\pm$ 0.013 & 2.31 $\pm$ 0.10 & 0.82 $\pm$ 0.03 & 0.5 $\pm$ 0.1 & 7.9 $\pm$ 0.2 \\
SPT2147-50 & -0.320 $\pm$ 0.009 & 0.339 $\pm$ 0.011 & 0.88 $\pm$ 0.05 & 0.212 $\pm$ 0.016 & 1.63 $\pm$ 0.23 & 0.72 $\pm$ 0.04 & 0.3 $\pm$ 2.4 & 4.8 $\pm$ 0.2 \\
\bottomrule
   \end{tabular}
   \tablecomments{\tablenotemark{a}\tablenotemark{b}Position relative to the lens. \tablenotemark{c}Total flux density. \tablenotemark{d}Major axis half-light radius. \tablenotemark{e}S\'ersic index. \tablenotemark{f}Axis ratio. \tablenotemark{g}Position angle. \tablenotemark{h}Magnification.}
   \label{tab:sourcepara}
\end{table*}

\begin{table*}
\scriptsize
   \centering
   \caption{Source parameters for single-channel CO models}
   \begin{tabular}{@{} lcccccccccc @{}}
\toprule
Source & $x_S$ & $y_S$ & $S$ & $a_S$ & $n_S$ & $b_S/a_S$ & $\phi_S$ & $\mu$ \\
 & (arcsec) & (arcsec) & (mJy) & (arcsec) &  &  & (deg CCW from E) &  \\
\midrule
SPT0346-52 & 0.228 $\pm$ 0.006 & 0.267 $\pm$ 0.009 & 2.25 $\pm$ 0.07 & 0.141 $\pm$ 0.005 & 0.48 $\pm$ 0.08 & 0.74 $\pm$ 0.05 & 10.6 $\pm$ 5.7 & 6.1 $\pm$ 0.1 \\
SPT0529-54 & 0.042 $\pm$ 0.011 & 0.022 $\pm$ 0.009 & 1.18 $\pm$ 0.26 & 0.597 $\pm$ 0.153 & 1.79 $\pm$ 0.34 & 0.45 $\pm$ 0.05 & 33.5 $\pm$ 3.5 & 10.4 $\pm$ 1.8 \\
SPT0532-50 & -0.006 $\pm$ 0.004 & 0.059 $\pm$ 0.003 & 1.70 $\pm$ 0.10 & 0.246 $\pm$ 0.018 & 1.12 $\pm$ 0.10 & 0.57 $\pm$ 0.03 & -18.7 $\pm$ 2.0 & 8.3 $\pm$ 0.4 \\
SPT2147-50 & -0.216 $\pm$ 0.015 & 0.327 $\pm$ 0.014 & 1.64 $\pm$ 0.29 & 0.355 $\pm$ 0.097 & 2.29 $\pm$ 0.70 & 0.72 $\pm$ 0.08 & 18.3 $\pm$ 10.4 & 5.2 $\pm$ 0.6 \\
\bottomrule
   \end{tabular}
   \label{tab:sourceparasingle}
\end{table*}

\begin{table*}
\scriptsize
   \centering
   \caption{Source parameters for SPT0346-52 CO(8-7) multiple-channel model}
   \begin{tabular}{@{} ccccccccc @{}}
\toprule
Velocity & $x_S$ & $y_S$ & $S$ & $a_S$ & $n_S$ & $b_S/a_S$ & $\phi_S$ & $\mu$ \\
(km/s) & (arcsec) & (arcsec) & (mJy) & (arcsec) &  &  & (deg CCW from E) &  \\
\midrule
-450 & 0.270 $\pm$ 0.014 & 0.366 $\pm$ 0.012 & 1.74 $\pm$ 0.09 & 0.072 $\pm$ 0.006 & 0.97 $\pm$ 0.09 & 0.71 $\pm$ 0.07 & 9.1 $\pm$ 13.2 & 3.3 $\pm$ 0.1 \\
-300 & 0.254 $\pm$ 0.012 & 0.331 $\pm$ 0.010 & 2.65 $\pm$ 0.13 & 0.119 $\pm$ 0.008 & 1.18 $\pm$ 0.07 & 0.89 $\pm$ 0.07 & 41.8 $\pm$ 11.8 & 4.4 $\pm$ 0.2 \\
-150 & 0.210 $\pm$ 0.008 & 0.273 $\pm$ 0.007 & 2.33 $\pm$ 0.11 & 0.154 $\pm$ 0.008 & 0.60 $\pm$ 0.06 & 0.66 $\pm$ 0.04 & 9.2 $\pm$ 5.0 & 6.4 $\pm$ 0.2 \\
0 & 0.182 $\pm$ 0.007 & 0.235 $\pm$ 0.006 & 2.40 $\pm$ 0.10 & 0.140 $\pm$ 0.007 & 0.76 $\pm$ 0.08 & 0.66 $\pm$ 0.05 & 10.1 $\pm$ 9.4 & 7.5 $\pm$ 0.2 \\
150 & 0.200 $\pm$ 0.007 & 0.227 $\pm$ 0.007 & 2.72 $\pm$ 0.12 & 0.132 $\pm$ 0.009 & 0.81 $\pm$ 0.08 & 0.71 $\pm$ 0.05 & 14.3 $\pm$ 11.9 & 7.2 $\pm$ 0.2 \\
300 & 0.201 $\pm$ 0.011 & 0.244 $\pm$ 0.008 & 1.62 $\pm$ 0.11 & 0.118 $\pm$ 0.012 & 0.97 $\pm$ 0.09 & 0.80 $\pm$ 0.05 & 37.7 $\pm$ 13.6 & 6.7 $\pm$ 0.3 \\
450 & 0.190 $\pm$ 0.019 & 0.259 $\pm$ 0.013 & 0.56 $\pm$ 0.04 & 0.118 $\pm$ 0.011 & 0.56 $\pm$ 0.04 & 0.85 $\pm$ 0.07 & 27.4 $\pm$ 9.9 & 7.0 $\pm$ 0.6 \\
\bottomrule
   \end{tabular}
   \label{tab:sourceparaSPT0346}
\end{table*}

\begin{table*}
\scriptsize
   \centering
   \caption{Source parameters for SPT0529-54 CO(6-5) multiple-channel model}
   \begin{tabular}{@{} ccccccccc @{}}
\toprule
Velocity & $x_S$ & $y_S$ & $S$ & $a_S$ & $n_S$ & $b_S/a_S$ & $\phi_S$ & $\mu$ \\
(km/s) & (arcsec) & (arcsec) & (mJy) & (arcsec) &  &  & (deg CCW from E) &  \\
\midrule
-300 & -0.032 $\pm$ 0.027 & 0.042 $\pm$ 0.017 & 0.81 $\pm$ 0.13 & 0.812 $\pm$ 0.110 & 1.29 $\pm$ 0.28 & 0.23 $\pm$ 0.06 & 40.7 $\pm$ 5.9 & 9.0 $\pm$ 1.0 \\
-50 & -0.003 $\pm$ 0.007 & -0.003 $\pm$ 0.008 & 1.01 $\pm$ 0.25 & 0.282 $\pm$ 0.070 & 1.58 $\pm$ 0.23 & 0.87 $\pm$ 0.08 & 49.5 $\pm$ 20.2 & 15.0 $\pm$ 2.6 \\
200 & 0.081 $\pm$ 0.013 & 0.048 $\pm$ 0.018 & 0.70 $\pm$ 0.16 & 0.341 $\pm$ 0.080 & 1.24 $\pm$ 0.29 & 0.35 $\pm$ 0.05 & 52.4 $\pm$ 4.5 & 15.0 $\pm$ 2.2 \\
450 & 0.031 $\pm$ 0.014 & -0.019 $\pm$ 0.014 & 0.67 $\pm$ 0.15 & 0.854 $\pm$ 0.109 & 1.34 $\pm$ 0.30 & 0.31 $\pm$ 0.13 & 47.4 $\pm$ 7.9 & 8.2 $\pm$ 1.1 \\
\bottomrule
   \end{tabular}
   \label{tab:sourceparaSPT0529}
\end{table*}

\begin{table*}
\scriptsize
   \centering
   \caption{Source parameters for SPT0532-50 CO(6-5) multiple-channel model}
   \begin{tabular}{@{} ccccccccc @{}}
\toprule
Velocity & $x_S$ & $y_S$ & $S$ & $a_S$ & $n_S$ & $b_S/a_S$ & $\phi_S$ & $\mu$ \\
(km/s) & (arcsec) & (arcsec) & (mJy) & (arcsec) &  &  & (deg CCW from E) &  \\
\midrule
-400 & 0.080 $\pm$ 0.006 & 0.049 $\pm$ 0.004 & 0.92 $\pm$ 0.06 & 0.117 $\pm$ 0.009 & 1.41 $\pm$ 0.14 & 0.65 $\pm$ 0.07 & -27.3 $\pm$ 8.0 & 9.0 $\pm$ 0.4 \\
-300 & 0.067 $\pm$ 0.004 & 0.059 $\pm$ 0.004 & 1.63 $\pm$ 0.07 & 0.122 $\pm$ 0.006 & 1.53 $\pm$ 0.14 & 0.75 $\pm$ 0.05 & -0.3 $\pm$ 0.1 & 9.6 $\pm$ 0.3 \\
-200 & 0.042 $\pm$ 0.003 & 0.058 $\pm$ 0.002 & 1.86 $\pm$ 0.12 & 0.129 $\pm$ 0.010 & 1.50 $\pm$ 0.10 & 0.73 $\pm$ 0.05 & -2.0 $\pm$ 1.6 & 10.7 $\pm$ 0.5 \\
-100 & 0.004 $\pm$ 0.000 & 0.048 $\pm$ 0.003 & 2.10 $\pm$ 0.11 & 0.150 $\pm$ 0.010 & 1.16 $\pm$ 0.09 & 0.82 $\pm$ 0.06 & -62.9 $\pm$ 10.3 & 11.2 $\pm$ 0.4 \\
0 & -0.024 $\pm$ 0.002 & 0.058 $\pm$ 0.005 & 2.22 $\pm$ 0.12 & 0.184 $\pm$ 0.011 & 0.63 $\pm$ 0.06 & 0.60 $\pm$ 0.04 & 117.3 $\pm$ 5.4 & 10.9 $\pm$ 0.5 \\
100 & -0.066 $\pm$ 0.005 & 0.089 $\pm$ 0.006 & 2.23 $\pm$ 0.12 & 0.173 $\pm$ 0.009 & 0.86 $\pm$ 0.07 & 0.82 $\pm$ 0.05 & -71.4 $\pm$ 11.7 & 10.1 $\pm$ 0.4 \\
200 & -0.092 $\pm$ 0.004 & 0.078 $\pm$ 0.004 & 1.73 $\pm$ 0.11 & 0.153 $\pm$ 0.011 & 1.33 $\pm$ 0.09 & 0.87 $\pm$ 0.07 & -66.6 $\pm$ 15.0 & 10.4 $\pm$ 0.5 \\
300 & -0.129 $\pm$ 0.010 & 0.055 $\pm$ 0.004 & 1.39 $\pm$ 0.08 & 0.141 $\pm$ 0.011 & 0.94 $\pm$ 0.07 & 0.77 $\pm$ 0.07 & 12.1 $\pm$ 2.5 & 10.2 $\pm$ 0.5 \\
400 & -0.131 $\pm$ 0.008 & 0.076 $\pm$ 0.005 & 0.68 $\pm$ 0.05 & 0.096 $\pm$ 0.010 & 1.42 $\pm$ 0.19 & 0.56 $\pm$ 0.05 & -21.1 $\pm$ 12.4 & 10.7 $\pm$ 0.6 \\
\bottomrule
   \end{tabular}
   \label{tab:sourceparaSPT0532}
\end{table*}

\begin{table*}
\scriptsize
   \centering
   \caption{Source parameters for SPT2147-50 CO(6-5) 75~km~s$^{-1}$-channel  model}
   \begin{tabular}{@{} ccccccccc @{}}
\toprule
Velocity & $x_S$ & $y_S$ & $S$ & $a_S$ & $n_S$ & $b_S/a_S$ & $\phi_S$ & $\mu$ \\
(km/s) & (arcsec) & (arcsec) & (mJy) & (arcsec) &  &  & (deg CCW from E) &  \\
\midrule
-300 & -0.446 $\pm$ 0.018 & 0.237 $\pm$ 0.013 & 2.75 $\pm$ 0.27 & 0.294 $\pm$ 0.048 & 2.44 $\pm$ 0.32 & 0.62 $\pm$ 0.07 & 10.3 $\pm$ 2.9 & 4.1 $\pm$ 0.2 \\
-225 & -0.419 $\pm$ 0.019 & 0.267 $\pm$ 0.013 & 2.50 $\pm$ 0.19 & 0.251 $\pm$ 0.032 & 2.66 $\pm$ 0.42 & 0.61 $\pm$ 0.07 & 0.1 $\pm$ 2.5 & 4.2 $\pm$ 0.2 \\
-150 & -0.382 $\pm$ 0.020 & 0.314 $\pm$ 0.017 & 2.74 $\pm$ 0.29 & 0.291 $\pm$ 0.042 & 2.19 $\pm$ 0.37 & 0.73 $\pm$ 0.08 & 4.3 $\pm$ 2.3 & 4.2 $\pm$ 0.2 \\
-75 & -0.284 $\pm$ 0.015 & 0.290 $\pm$ 0.012 & 3.37 $\pm$ 0.31 & 0.327 $\pm$ 0.040 & 2.08 $\pm$ 0.32 & 0.74 $\pm$ 0.08 & 2.1 $\pm$ 2.4 & 4.9 $\pm$ 0.3 \\
0 & -0.212 $\pm$ 0.011 & 0.298 $\pm$ 0.012 & 3.54 $\pm$ 0.29 & 0.325 $\pm$ 0.040 & 2.61 $\pm$ 0.32 & 0.67 $\pm$ 0.07 & -0.7 $\pm$ 2.2 & 5.2 $\pm$ 0.3 \\
75 & -0.189 $\pm$ 0.010 & 0.375 $\pm$ 0.015 & 3.14 $\pm$ 0.28 & 0.263 $\pm$ 0.036 & 2.00 $\pm$ 0.29 & 0.73 $\pm$ 0.08 & 1.7 $\pm$ 2.2 & 5.5 $\pm$ 0.3 \\
150 & -0.191 $\pm$ 0.014 & 0.430 $\pm$ 0.017 & 2.86 $\pm$ 0.33 & 0.357 $\pm$ 0.058 & 1.67 $\pm$ 0.33 & 0.43 $\pm$ 0.07 & -1.8 $\pm$ 1.9 & 4.9 $\pm$ 0.3 \\
225 & -0.153 $\pm$ 0.034 & 0.443 $\pm$ 0.024 & 1.05 $\pm$ 0.19 & 0.324 $\pm$ 0.054 & 0.98 $\pm$ 0.53 & 0.43 $\pm$ 0.09 & 2.4 $\pm$ 2.3 & 5.3 $\pm$ 0.5 \\
\bottomrule
   \end{tabular}
   \label{tab:sourceparaSPT2147dv75}
\end{table*}

\begin{table*}
\scriptsize
   \centering
   \caption{Source parameters for SPT2147-50 CO(6-5) 150~km~s$^{-1}$-channel model}
   \begin{tabular}{@{} ccccccccc @{}}
\toprule
Velocity & $x_S$ & $y_S$ & $S$ & $a_S$ & $n_S$ & $b_S/a_S$ & $\phi_S$ & $\mu$ \\
(km/s) & (arcsec) & (arcsec) & (mJy) & (arcsec) &  &  & (deg CCW from E) &  \\
\midrule
-262.5 & -0.415 $\pm$ 0.018 & 0.249 $\pm$ 0.010 & 2.52 $\pm$ 0.29 & 0.259 $\pm$ 0.050 & 2.38 $\pm$ 0.42 & 0.70 $\pm$ 0.09 & 4.0 $\pm$ 2.2 & 4.3 $\pm$ 0.2 \\
-112.5 & -0.331 $\pm$ 0.015 & 0.313 $\pm$ 0.013 & 3.49 $\pm$ 0.38 & 0.407 $\pm$ 0.063 & 3.02 $\pm$ 0.45 & 0.73 $\pm$ 0.07 & -3.0 $\pm$ 3.6 & 4.0 $\pm$ 0.3 \\
37.5 & -0.215 $\pm$ 0.009 & 0.354 $\pm$ 0.014 & 3.49 $\pm$ 0.32 & 0.313 $\pm$ 0.041 & 2.02 $\pm$ 0.26 & 0.82 $\pm$ 0.06 & -1.2 $\pm$ 3.3 & 5.2 $\pm$ 0.3 \\
\bottomrule
   \end{tabular}
   \label{tab:sourceparaSPT2147dv150}
\end{table*}
\bibliography{DMS_CO_aastex.bib}
\end{CJK*}
\end{document}